\documentclass[useAMS,usenatbib,referee]{mn2e}

\usepackage{epsf}

\newcommand{\bmth}[1]{\mbox{\boldmath${#1}$}}

\title[Dynamic tides in rotating objects]
{Dynamic tides in rotating objects: orbital
circularisation of extra solar planets 
for realistic planet models}
\author[P. B. Ivanov and
J. C. B. Papaloizou]{
P. B. Ivanov$^{1,2}$\thanks{E-mail:P.Ivanov@damtp.cam.ac.uk
(PBI) J.C.B.Papaloizou@damtp.cam.ac.uk (JCBP)} 
and J. C. B. Papaloizou$^{1}$
\footnotemark[1]\\
$^{1}$Department of Applied Mathematics and Theoretical Physics,
University of Cambridge,\\
Centre for Mathematical Sciences, 
Wilberforce Road, Cambridge, CB3 0WA, UK \\
$^{2}$Astro Space Centre, P. N. Lebedev Physical Institute,
4/32 Profsoyuznaya Street,
Moscow, 117810, Russia}

\begin{document}

\date{Accepted Received ; in original form }

\pagerange{\pageref{firstpage}--\pageref{lastpage}} \pubyear{2002}

\maketitle

\label{firstpage}

\begin{abstract}
We consider the general problem of the tidal capture or circularisation
from large eccentricity of a uniformly rotating object.
We extend the self-adjoint formalism introduced 
in our recent paper(Papaloizou \& Ivanov 2005 hereafter PI))  to  
derive  general expressions for the energy and angular momentum transfered 
 when  the  planet or a star
 passes through  periastron in a parabolic or highly eccentric 
orbit around a central mass. These
can be used  without making a low frequency approximation as was done in PI.
We show how these can be adapted  to the low frequency limit in which only inertial modes contribute
for baratropic planet models. 
In order to make quantitative estimates, we  calculate the inertial mode eigenspectrum  for 
planet models of  one and five  Jupiter masses $M_J,$  without a solid core, with different
radii corresponding to different ages. 
The spectra are found in general to be more complex than of a polytrope with index  $n=1.5,$ considered in PI, 
because of the existence  global modes associated with the transition from molecular
to metallic Hydrogen. Nonetheless the main tidal response is still found to
be determined  by two global modes which have polytropic counterparts.

These also determine the uniform angular velocity in a state of pseudo synchronisation, for which
the angular momentum transfered during an encounter is zero. This is found
to be close to $1.55$  times the circular orbit angular velocity at periastron
for all models considered. This is in contrast to the situation
when only the $f$ mode is considered (Ivanov \& Papaloizou 2004, hereafter IP)
and the equilibrium angular velocity is found to be much larger.

We consider the multi-passage problem when there is no dissipation
finding that stochastic instability  resulting 
in the  stochastic gain of inertial mode energy over many periastron passages
occurs under similar conditions to those already found by IP for the $f$ modes.
We find that this requires  circularisation to  start with a  semi-major axis
exceeding $\sim 30 AU,$ for final periods of $\sim 3$ days.
reducing to $\sim 1-2 AU$  for final
periods $\sim 1.2$ days.

Finally we  apply our  calculations of the energy transfer during a periastron passage
to  the problem of the  tidal circularisation of  
the orbits of the extrasolar planets in a state of pseudo
synchronisation, expected because of the relatively small
inertia of the planet,  and  find that inertial mode excitation
dominates the tidal interaction for $1 M_J$ planets that start with semi- major axes
less than $10 AU$  and   end up on circular orbits with final
period in the $4-6$ day range.  It is potentially 
able to account for initial circularisation
up to a final $6$ day  period within a few $Gyr$  But in the case of $5M_J$
oscillation modes excited in the star are more important.

\end{abstract}

\vspace{-1cm}
\section{Introduction}

The process of  tidal capture, whereby a body in  initially
unbound orbit has a close approach to another gravitating
mass, and through the excitation of internal modes
of oscillation, becomes subsequently bound is of general importance
in astrophysics. It is believed to play a role in binary formation
in globular clusters 
(eg. Press \& Teukolsky 1977, hereafter PT,  and references therein)
and to play a role in the tidal interactions of stars in galactic centres.
A  related problem is the circularisation of an orbit from large 
eccentricity which may occur after a tidal capture. In this  situation,
each periastron passage is very similar to a  close encounter in a weakly
unbound orbit  that causes a change in the energy contained
within internal modes of oscillation at the expense of that in the
orbit. 

In order to estimate
capture probabilities and circularisation time scales the excitation of 
normal modes of oscillation by a perturbing potential must be considered.
In general modes associated with spherical stars, or small perturbations
of them,  have been considered
(eg. PT, IP and references therein). However, if the tidally perturbed body has multiple
encounters, angular momentum transfer should lead it
to rotate at an angular velocity related to that at periastron.
In this case  modes of oscillation with periods comparable
to the rotation period need to be considered (eg. Papaloizou \& Pringle 1978).

An approach for dealing with the problem of calculating tidal phenomena
while taking into account the part of the oscillation spectrum with eigen frequencies
comparable to the rotation frequency was recently
proposed by PI. They adopted a self-adjoint
formalism for calculating the tidal response that used the density perturbation to describe 
the motions within the star and they made a low frequency approximation that the rotation
and mode frequencies were very much less than the inverse sound crossing time.
They also evaluated the inertial mode spectrum for a polytrope with index $n=3/2$
and showed that inertial modes dominated the response at low frequencies or large
periastron distances.

Given the importance of rotation, the purpose of this paper is to develop further 
the work begun in PI
and to use the results to  obtain time scales for tidal phenomena such as circularisation
of the orbit starting from high eccentricity.

We begin by noting that the self-adjoint formalism can be extended
to general linear displacements without making a low frequency approximation.
In this form, which uses the Lagrangian displacement to describe the perturbations
it has wider applicability than the formalism presented in PI which made a low frequency
approximation. We also wish to evaluate the expressions obtained for the
energy and angular momentum exchanged during close periastron passages for realistic planet models.
Accordingly we calculate sequences of  models  of one Jupiter mass $M_J$
and $5M_J.$ Each sequence  has a range of radii
in order to take account of the fact that the planet radius varies with age.
For simplicity we do not include a solid core noting that according to
Saumon \& Guillot (2004) observations of Jupiter are not inconsistent with this.
We calculate the oscillation  spectra together with required
overlap integrals for the models using a set of basis or trial functions.

We then use the determined eigenmode properties for these models
to estimate tidal phenomena such as the attainment of pseudo
synchronism for which the angular momentum transfer at periastron passage is zero.
Such a state is expected to be attained, on account of the small inertia
of the planet compared to that of the orbit, with subsequent energy exchanges occurring
with the planet rotating at an equilibrium angular velocity.
The inclusion of inertial modes is important for determining this angular velocity
which turns out to be  close to that of the orbit at periastron.
Calculations in IP incorporating only the  $f$ modes gave a much higher value.

Having calculated the equilibrium angular velocity we are able to consider multiple encounters
and determine conditions for the onset of stochastic instability, that  results 
in the  stochastic gain of inertial mode energy over many periastron passages when there is
no dissipation (Kochanek 1992, Mardling 1994a,b). Then
we make estimates of circularisation time scales
for the two different planet masses taking into account the evolutionary variation of the 
planet radius with age. 

The plan of the paper is as follows. In section 2 we formulate the new self-adjoint formalism 
for the problem of linear perturbations proposed in our recent paper(PI) using the Lagrangian 
displacement to describe the motions and then use it to  
derive  general expressions for the energy and angular momentum transfered 
when a rotating planet or a star passes through periastron in a parabolic or highly eccentric 
orbit around a central mass without making a low frequency approximation as in PI.
It can thus be applied to general situations in which $p,$ $f,$ $g$ and inertial
modes are significant. 
In section 3  we show that the general formalism can be reduced to that of PI
in an appropriate low frequency limit.  In this case only the inertial modes
contribute to the tidal response for the baratropic stellar models we consider.
In section 4 we discuss the tidal energy and angular momentum transfered through
inertial modes giving very simple expressions that apply when the planet or star rotates
in a state of pseudo synchronisation for which the angular momentum transfered
in an encounter is essentially zero.
We apply these to  the multi-passage problem when there is no dissipation
finding conditions for the occurrence of the  stochastic instability.  
We find that this requires the circularisation process to start with a rather
large semi-major axis $\sim 30 AU,$ for final periods of $\sim 3$ days.
However, this could be reduced to $\sim 1-2 AU$ for the shortest observed final
periods $\sim 1.2$ days.

In section 5  we  calculate the eigenspectrum of the inertial modes for the
planet models which have been obtained using
 a realistic equation of state.  We find that the counterparts of the two
global modes found in a polytrope with $n=1.5$ exist in these cases but that
there are additional global modes associated with the transition from molecular
to metallic Hydrogen.
We go on to apply the results
to  the problem of tidal circularisation of  
the orbits of the extrasolar planets in sections 6 and 7 finding that inertial modes
dominate the tidal interaction for $M_J$ planets ending up on circular orbits with final
period in the $4-6$ day range and are 
potentially  able to account for the initial circularisation
at a $6$ day  period within a few $Gyr$ in that case. However, in the case of $5M_J$
oscillation modes excited in the star become more important.
Finally in section 8 we discuss our results.

\section{Formal solution of the tidal problem}

In this Section we further develop the self adjoint formalism describing the
pulsations of  a uniformly 
rotating fully convective object (a planet or a star), 
referred to hereafter  as  a planet,
introduced in our previous Paper (Papaloizou $\&$ Ivanov 2005, PI).
 We give general expressions for the energy and angular momentum
transfer from the orbit to the planet as a result of a  close encounter or
fly by around 
a gravitating centre. Here, contrary to PI, we do  not  make the assumption
that only low frequency modes respond effectively to the tidal forcing
and the results obtained  are valid for the excitation of  any pulsational
mode of a rotating planet.

The fundamental quantity used in our analysis 
is the Lagrangian displacement vector ${\bmth{ \xi}}$. By definition,
this is real. We adopt a cylindrical coordinate system
$(\varpi, \phi, z)$ with origin at the centre of mass of the planet.

We make use of the
Fourier transforms of the displacement vector and any other perturbation 
quantity, say Q, over the azimuthal angle  $\phi$ and time $t$ in
the form
\begin{equation}
Q = \sum_{m}\left(  \exp({im\phi})\int^{+\infty}_{-\infty}d\sigma \tilde
Q_{m}\exp({-i\sigma t}) + cc \right ), \label{eq p1} \end{equation}
where the sum is over $m=0$  and $2$ and $cc$ denotes
the complex conjugate. The reality of $Q$
implies that the   Fourier transform, indicated by tilde  satisfies 
$\tilde Q_{m}(\sigma) = \tilde
Q_{-m}^*(-\sigma).$

The inner products of two complex
scalars $Y_{1}$, $Y_{2}$, and
vectors ${\bmth{ \eta}_{1}}$ and ${\bmth{ \eta}_{2}}$ are defined as
\begin{equation}
(Y_{1}|Y_{2})=\int \varpi d\varpi dz\rho 
(Y_{1}^{*}Y_{2})
\quad
({\bmth{ \eta}}_{1}|{\bmth{ \eta}}_{2})=\int \varpi d\varpi dz\rho 
({\bmth{ \eta}}_{1}^{*}\cdot {\bmth{ \eta}}_{2}), \label{eq p2} 
\end{equation}
where $({\bmth{ \eta}}^{*}\cdot {\bmth{ \mu}})$ is the scalar
product, $\rho$ is the density, and $*$ stands for 
the complex conjugate. 

\subsection{Perturbed equations of motion}

We assume that, viewed from an inertial frame,
the planet is rotating with uniform angular velocity
$\bmth {\Omega }$ directed perpendicular to the orbital plane. 
In this  situation the
hydrodynamic equations for the perturbed quantities 
take the simplest form in the rotating
frame which we accordingly adopt.

Since the planet is fully convective,
the entropy per unit of mass of the planetary gas remains
approximately the same over the volume of the planet
\footnote{Note that this condition can be broken at the
radius corresponding to phase transition of hydrogen from
metallic to molecular form, see below.}, and the pressure
$P$ can be considered as a function of density $\rho$ only,
$P=P(\rho)$. A perturbation of the planet may be
considered as adiabatic, and therefore 
the same functional dependence 
$P=P(\rho)$ holds during  perturbation as well. 
This condition 
is often referred to as the baratropic condition.  
In the  baratropic approximation the linearised 
Navier-Stokes equations take the form (see PI)
\begin{equation}
{D^{2} {\bmth{ \xi}} \over Dt^{2}}+2{\bmth {\Omega}}\times {D \bmth
{\xi}\over Dt}=-\nabla W-{\bmth {f}}_{\nu}, \label{eq p3} 
\end{equation}
where
\begin{equation}
W=c_{s}^{2}\rho^{'}/\rho+\Psi^{int}+\Psi^{ext},  \label{eq p4}
\end{equation}
$\rho^{'} $ is the
density perturbation, $c_{s}$ is the 
adiabatic sound speed, $\Psi^{int}$ is the  stellar  gravitational 
potential  arising from the  perturbations and  $\Psi^{ext}$ is 
the external forcing  tidal potential. The 
viscous force per unit of mass is  ${\bmth {f}}_{\nu}.$ In
Cartesian coordinates  it can be written in component form as
\begin{equation}
f^{\alpha}_{\nu}=-(1/\rho)t^{\alpha, \beta}_{,\beta}=
(1/\rho)\lbrace \rho \nu (\dot \xi^{\alpha}_{,\beta}+
\dot \xi^{\beta}_{,\alpha}-
{2\over 3}\dot
\xi^{\gamma}_{,\gamma}\delta^{\beta}_{\alpha})\rbrace_{,\beta},
\label{eq p5}
\end{equation} 
where the comma denotes differentiation and $\nu$ is the kinematic
viscosity. The Greek indices enumerate spatial coordinate directions 
and the usual summation convention over these  is adopted.

Note that the centrifugal term is absent in equation  $(\ref{eq p3})$
being formally incorporated into the potential governing
the static equilibrium of the unperturbed star.
The convective derivative ${D\over Dt} \equiv {\partial \over \partial t}$
as there is no unperturbed motion in the rotating frame.

\noindent The linearised continuity equation is    
\begin{equation}
\rho^{'}=-\nabla \cdot (\rho {\bmth {\xi}}),  \label{eq p6}
\end{equation}
and the linearised Poisson equation is
\begin{equation}
\Delta \Psi^{int}=4\pi
G\rho^{'} \label{eq p7}.
\end{equation}

Provided that  expressions for the density, sound speed and
kinematic viscosity are specified in some unperturbed model of
the planet, the set of equations  $(\ref{eq p3}-\ref{eq p7})$ is complete.
Without the
viscous term on the right hand side of equation $(\ref{eq p3})$ these equations
coincide with the analogous equations used by PI.

The perturbed potential $\Psi^{int}$ can be formally expressed in
terms of the density perturbation with help of the inverse of the 
Laplace operator, $\Delta^{-1}$, and the density perturbation is
expressed in terms of the displacement vector with help of the
continuity equation. We have
\begin{equation}
\Psi^{int}=-4\pi G \Delta^{-1} \nabla \cdot (\rho {\bmth {\xi}}). 
\label{eq p8}
\end{equation} 

Now we express $W$ in terms of ${\bmth {\xi}}$ with help of 
equations $(\ref{eq p4})$, $(\ref{eq p6})$ and $(\ref{eq p8})$.
We substitute the result into equation  $(\ref{eq p3})$ thus obtaining an 
equation containing only the displacement vector ${\bmth {\xi}}$ as
a dynamical variable. After making Fourier transform of this equation
we get
\begin{equation}
\sigma^{2}\tilde {\bmth {\xi}}_{m}-\sigma {\bmth {\cal B}}\tilde 
{\bmth {\xi}}_{m} -{\bmth {\cal C}}\tilde {\bmth {\xi}}_{m}-
{\bmth {f}}_{\nu}(\tilde {\bmth {\xi}}_{m})  =
\tilde {\bmth S}_{m}, \label{eq p9}
\end{equation}
where the operators ${\bmth {\cal B}}$ and ${\bmth {\cal C}}$ are
defined through their action on the vector ${\bmth {\xi}}$ as
\begin{equation}
{\bmth {\cal B}} {\bmth {\xi}} =-2i{\bmth {\Omega}}\times {\bmth{\xi}},
\label{eq p10}
\end{equation} 
and
\begin{equation}
{\bmth {\cal C}} {\bmth {\xi}}=-\nabla (({c_{s}^{2}\over \rho}+4\pi G\Delta^{-1})
\nabla \cdot (\rho {\bmth {\xi}}))
\label{eq p11}
\end{equation}
It can be easily checked that the so defined operators 
${\bmth {\cal B}}$ and ${\bmth {\cal C}}$ are self adjoint in the
sense of the normal product $(\ref{eq p2})$. Also for the models we
consider with weak self-gravity, the operator 
 ${\bmth {\cal C}}$  may be taken to be positive definite for displacements
with  a  non zero density perturbation somewhere and non negative
for arbitrary displacements including those with zero density perturbation everywhere.

The source term  $\tilde {\bmth S}_{m}$ is determined by the external 
forcing potential
\begin{equation}
\tilde {\bmth {S}}_{m}=\nabla \tilde \Psi^{ext}_{m}.
\label{eq p12}
\end{equation}

\subsection{Self adjoint form of the perturbed equation of motion and
its formal solution}

The presence of the viscous force ${\bmth {f}}_{\nu}$ leads to 
dissipation of energy
of the perturbed motion on a dissipation time scale.  For global disturbances 
in  
astrophysical systems this time scale is typically very large compared to 
a time scale characterising the tidal forcing. In this situation we
can treat the action of the viscous force as a perturbation.

Neglecting the contribution of the viscous term,  
we see equation $(\ref{eq p9})$ has 
an  analogous form  to  one discussed by PI for low
frequency perturbations.   There it was  shown that
 this form led to  
a standard type of  eigen value problem  associated with a new 
self adjoint operator  ${\bmth {\cal H}}$. In order to bring equation
$(\ref{eq p9})$ to this standard  type, let us  formally consider the 
square root of the operator  ${\bmth {\cal C}}$, 
${\bmth {\cal C }}^{1/2}$, defined by the condition:
${\bmth {\cal C }}={\bmth {\cal C }}^{1/2}{\bmth {\cal C }}^{1/2}$
This can be done on account of the non negativity of ${\bmth {\cal C }}$. 
Further, let us introduce a  six dimensional vector $\vec Z$ with components
\begin{equation} 
{\bmth {Z}}_{1}=\sigma \tilde {\bmth {\xi}}_{m} ,
\quad {\bmth {Z}}_{2}={\bmth {\cal C}}^{1/2}\tilde {\bmth{ \xi}}_{m}. 
\label{eq p13}
\end{equation}  
It follows from
equation $(\ref{eq p9})$ that   the eigenvalue problem can 
be written in the standard form
\begin{equation}
\sigma \vec Z ={\bmth {\cal H}} \vec Z +\vec S,  \label{eq p14} 
\end{equation}
where the operator ${\bmth {\cal H}}$ has a matrix structure
\begin{equation}
{\bmth {\cal H}}=  
\left( \begin{array}{cc} {\bmth {\cal B}}
& 
{\bmth {\cal C}}^{1/2} \\
{\bmth {\cal C}}^{1/2} & 0 \end{array}\right), \label{eq p15}
\end{equation}  
and the source six dimensional vector $\vec S$ has the components
\begin{equation}
{\bmth{S}}_{1}=\tilde {\bmth { S}}_{m}, \quad {\bmth{S}}_{2}=0.
\label{eq p16}
\end{equation}
It follows from equation  $(\ref{eq p15})$  that the operator 
${\bmth {\cal H}}$
has the required self adjoint form. Now we can look for solution to 
equation  $(\ref{eq p14})$ in the form:
\begin{equation}
\vec Z=\sum_{k} \alpha_{k} \vec Z^{k}, \label{eq p17}
\end{equation} 
where 
$\vec Z_{k}$ are the eigenfunctions of ${\bmth{\cal H}}$ which satisfy
\begin{equation}  
\sigma_{k} \vec Z^{k} ={\bmth{\cal H}} \vec Z^{k}.  \label{eq p18} 
\end{equation}
The eigen frequencies $\sigma_k$ are necessarily real. Equation
$(\ref{eq p18})$ is equivalent to equation $(\ref{eq p9})$ with
the source term $\tilde {\bmth S}_{m}=0$. 

Note that the eigen spectrum of ${\bmth{\cal H}}$ or
a part of it can be continuous. It this case the summation in
$(\ref{eq p18})$ must be replaced by integration with an appropriate
measure. 

The eigen functions $\vec Z^{k}$ are orthogonal in the sense of the
inner product
\begin{equation}
<\vec Z^{k}| \vec Z^{l}>=({\bmth {Z}}^{k}_{1}|{\bmth {Z}}^{l}_{1})+
({\bmth {Z}}^{k}_{2}|{\bmth {Z}}^{l}_{2})=
\sigma_{k}\sigma_{l}({\bmth {\xi}}_{k}|
{\bmth {\xi}}_{l})+
({\bmth {\xi}}_{k}|{\bmth {\cal C}}{\bmth
{\xi}}_{l}), \label{eq p19}
\end{equation}  
where ${\bmth {\xi}}_{k}={\bmth {Z}}_{1}^{k}/\sigma_{k}$

Substituting $(\ref{eq p17})$ into equation $(\ref{eq p14})$ and using 
$(\ref{eq p19})$ we obtain
\begin{equation}  
\alpha_{k} ={<\vec Z_{k}| \vec S>\over 
N_{k}(\sigma+i\sigma_{\nu}-\sigma_{k})}=  
{\sigma_{k}^{2}S_{k}\over 
N_{k}(\sigma+i\sigma_{\nu}-\sigma_{k})},  
\label{eq p21}
\end{equation}
where we add a small imaginary part to the frequency $\sigma$ with
$\sigma_{\nu} > 0$  in potentially resonant denominators
to account for dissipation. As we show below, for modes for which
the effect of viscosity can be treated as a small perturbation, 
an explicit expression for $\sigma_{\nu}$ can be obtained by taking
into account the dissipation determined by the viscous force 
${\bmth {f}}_{\nu}$.
\begin{equation}
N_{k}=<\vec Z_{k}| \vec Z_{k}>=\sigma_{k}^{2}({\bmth {\xi}}_{k}|
{\bmth {\xi}}_{k})+({\bmth {\xi}}_{k}|{\bmth {\cal C}}{\bmth{\xi}}_{k})          
\label{eq p21a}
\end{equation} 
is the norm.

The coefficients $S_{k}$ determine decomposition of the source vector
over the eigen functions, 
\begin{equation}
\tilde {\bmth {S}}_{m}=\sum_{k}{<\vec Z_{k}|\vec S>\over N_{k}}{\bmth {Z}}_{1}=
\sum_{k} {\sigma_{k}^{2}S_{k}\over N_{k}}{\bmth {\xi}}_{k},
\label{eq p22a}
\end{equation}
where summation is performed over all modes with a particular value of
$m$. They can be written in the form
\begin{equation}
S_{k}=( {\bmth {\xi}}_{k}| \tilde {\bmth {S}}_{m})=
(\rho^{'}_{k}|\tilde \Psi_{m}^{ext}),
\label{eq p22} 
\end{equation}
where $\rho_{k}'$ is the density perturbation associated with the 
eigenvector ${\bmth {\xi}}_{k}$: $\rho_{k}' =-\nabla \rho {\bmth
{\xi}}_{k}$, and  
to obtain the last equality, we integrate by parts.

It is important to note that the coefficients $S_{k}$, the eigen
frequencies $\sigma_{k}$ and the  associated displacement vector 
$\tilde {\bmth {\xi}}_{k}$
obey a certain identity relation
\begin{equation}
0=\sum_{k} {\sigma_{k}S_{k}\over N_{k}} {\bmth {\xi}}_{k}, \label{eq p23}  
\end{equation}   
which follows from equations $(\ref{eq p13})$, 
$(\ref{eq p22a})$
and the orthogonality relation
$(\ref{eq p19})$ for the eigen vectors. This relation is also required to ensure
that $ \tilde {\bmth {\xi}}_{m}$ remains bounded as $\sigma \rightarrow 0$
(see below).

The solution for the displacement vector $\tilde {\bmth {\xi}}_{m}$
follows from equations $(\ref{eq p13})$, $(\ref{eq p17})$ 
and $(\ref{eq p21})$
\begin{equation}
\tilde {\bmth {\xi}}_{m} =
\sum_{k} {\sigma_{k}^{2} S_{k}
\over N_{k} \sigma (\sigma+i\sigma_{\nu}-\sigma_{k})}{\bmth {\xi}}_{k}=
\sum_{k} {\sigma_{k} S_{k}
\over N_{k}(\sigma+i\sigma_{\nu}-\sigma_{k})}{\bmth {\xi}}_{k},
\label{eq p24}  
\end{equation}
where we use the relation $(\ref{eq p23})$ to obtain the last
equality.

The displacement vector ${\bmth {\xi}}$ is obtained from its Fourier's
transform $\tilde {\bmth {\xi}}_{m}$ by integration over $\sigma $ and
summation over $m$
\begin{equation}
{\bmth {\xi}}=\sum_{m,k}\int d\sigma \lbrace {\sigma_{k} S_{k}
\over N_{k}(\sigma+i\sigma_{\nu}-\sigma_{k})}{\bmth {\xi}}_{k}
e^{-i\sigma t}e^{im\phi}+c.c. \rbrace .
\label{eq p25} 
\end{equation}
This expression can be simplified in the limit $t \rightarrow \infty$ 
 at which only the poles in the expression in the braces contribute 
significantly to the integral over $\sigma $. In this limit we may,
therefore, write
\begin{equation}
{\bmth {\xi}}=2\pi i\sum_{m,k} ({\sigma_{k} S_{k}\over N_{k}}
e^{-\sigma_{\nu}t-i\sigma_{k}t}e^{im\phi}{\bmth {\xi}}_{k}+c.c.).
\label{eq p26} 
\end{equation}

\subsection{Energy and angular momentum transfer}

It can be easily shown from equation $(\ref{eq p3})$ or equation
$(\ref{eq p9})$ that the expression for the canonical energy of the
perturbations is given by
\begin{equation}
E_{c}=\int_{V}d^{3}x \rho {1\over 2}({\dot {\bmth {\xi}}}^{2}+
{\bmth {\xi}}\cdot {\bmth {\cal C}}{\bmth {\xi}})
= \pi \sum_{m} \lbrace (\dot {\bmth {\xi}}_{m}| \dot {\bmth {\xi}}_{m})+
({\bmth {\xi}}_{m}| {\bmth {\cal C}}{\bmth {\xi}}_{m})\rbrace,
\label{eq p27} 
\end{equation}
where we integrate over the volume of the planet. 

We evaluate the expression $(\ref{eq p27})$ after a close encounter,
 $t > 0$. Assuming that the time  after 
  closest approach is much larger than  the characteristic time scale of the 
 encounter, but nevertheless  much smaller than 
the characteristic viscous time scale $\sim 1/\sigma_{\nu}$ , 
we can use equation $(\ref{eq p26})$
setting $\sigma_{\nu}=0.$    Substituting the expression for 
${\bmth {\xi}}$ and its time derivative into $(\ref{eq p27})$ and 
taking into account equation $(\ref{eq p21a})$
we get
\begin{equation}
E_{c}=8\pi^{3}\sum_{m,k}{\sigma_{k}^{2}S_{k}S_{k}^{*}\over N_{k}},
\label{eq p28}
\end{equation}
where we use the orthogonality relation $(\ref{eq p19})$ and the
expression for the norm $(\ref{eq p19})$. 

The rate of viscous dissipation of energy is given by 
the standard expression
\begin{equation}
\dot E_{\nu}=\int_{V}d^{3}x\rho \nu 
\lbrace \dot \xi^{\alpha}_{,\beta}(\xi^{\alpha}_{, \beta}+\xi^{\beta}_{, \alpha})
-{2\over 3}{(\dot \xi^{\alpha}_{, \alpha})}^{2} \rbrace.
\label{eq p29}
\end{equation}

An explicit expression for the amount of energy dissipated by viscosity
after the fly by, $\Delta E_{\nu} =\int^{+\infty}_{-\infty}dt \dot
E_{\nu}$, can be obtained from $(\ref{eq p29})$  substituting 
$(\ref{eq p25})$ into $(\ref{eq p29})$ and integrating over time 
the resulting expression. We have
\begin{equation}
\Delta E_{\nu} =8\pi^{2}\int d\sigma \sigma^{2} \sum_{m,k,l}
{\sigma_{k}\sigma_{l}S_{k}\over N_{k}N_{l}(\sigma
-\sigma_{k}+i\sigma_{\nu})(\sigma
-\sigma_{l}-i\sigma_{\nu})}S_{k}S^{*}_{l}a_{k,l},
\label{eq p30}
\end{equation}
where we use the well known relation $\int dt e^{i(\sigma
-\sigma^{'})}=2\pi \delta (\sigma -\sigma^{'})$, and
\begin{equation}
a_{k,l}=\int \varpi d\varpi dz 
\lbrace \xi^{\alpha}_{\beta,k}((\xi^{\alpha}_{\beta, l})^{*}+(\xi^{\beta}_{\alpha, l})^{*})
-{2\over 3}\xi^{\alpha}_{\alpha, k}(\xi^{\beta}_{\beta, l})^{*} \rbrace.
\label{eq p31}
\end{equation}
As we have mentioned above,  we assume that viscosity is very
small, and the imaginary part of the frequency is much
smaller than the corresponding real part. In this case, one
can  see from equation $(\ref{eq p30})$ that the leading contribution
to $\Delta E_{\nu}$ is determined by diagonal terms in the summation
series with $k=l$. Also, only a part of the integral over $\sigma $
very close to the resonance, $\sigma \sim \sigma_{k}$, gives a
significant contribution. We obtain
\begin{equation}
\Delta E_{\nu}=8\pi^{2}\sum_{m,k}{\sigma_{k}^{4}S_{k}S^{*}_{k}
\over N_{k}^{2}}a_{k,k}
\int \left({d\sigma \over (\sigma - \sigma_{k})^{2}+\sigma_{\nu}^{2}}\right)=
{8\pi^{3}\over \sigma_{\nu}}\sum_{m,k}
{\sigma_{k}^{4}S_{k}S^{*}_{k}\over N_{k}^{2}}a_{k,k}.
\label{eq p32}
\end{equation}
The amount of energy transfered by tides into the modes must be 
equal to the amount of energy dissipated by viscosity: $E_{c}=\Delta
E_{\nu}$. From this condition and equation $(\ref{eq p28})$ and 
$(\ref{eq p32})$ we easily obtain an explicit expression 
for the inverse dissipation time 
\begin{equation}  
\sigma_{\nu}={\sigma_{k}^{2}a_{k,k}\over N_{k}}.
\label{eq p33}
\end{equation}

Similar arguments can be used for calculations of the amount of 
the angular momentum transferred to the modes after the fly by. 
However, the expression for the angular momentum transfer directly 
follows from the general relation between canonical energy and 
angular momentum corresponding to a particular mode  with given 
values of $k$ and $m$  (eg. Friedman and Schutz 1977) 
\begin{equation}
L_{c}=mE_{c}/\sigma_{k}.
\label{eq p34}
\end{equation} 
Accordingly, we have from equation $(\ref{eq p28})$   
\begin{equation}
L_{c}=16\pi^{3}\sum_{2,k}\sigma_{k}{S_{k}S_{k}^{*}\over N_{k}}.
\label{eq p35}
\end{equation}

The expression for the energy of the modes in the inertial frame, $E_{I}$,
follows directly from equations  $(\ref{eq p28})$ and $(\ref{eq p35})$
\begin{equation}   
E_{I}=E_{c}+\Omega L_{c}=
8\pi^{3}\left\lbrace \sum_{2,k}
\left(\sigma_{k}(\sigma_{k}+2\Omega){S_{k}S_{k}^{*}\over N_{k}}\right) 
+\sum_{0,k}\sigma^{2}_{k}{S_{k}S_{k}^{*}\over N_{k}}\right\rbrace .
\label{eq p37}
\end{equation}
Note that the modal energy in the inertial frame can be negative.
\section{Response of low frequency modes to the tidal forcing}

The formalism developed in the previous Section can be applied for any
pulsational mode of a rotating planet. It follows from equations $(\ref{eq p22})$,
 $(\ref{eq p28})$ and  $(\ref{eq p35})$ that in order to find the transfer 
of energy and angular momentum from the orbit to the modes we should 
be able to find solutions (eigen functions and eigen frequencies )
of equation  $(\ref{eq p15})$ describing free oscillations of the planet.
This can be done numerically.
However, in general, it is rather difficult 
to use numerical methods directly in solving equation $(\ref{eq p15})$ 
without any further simplifications. 
It turns out that these simplifications can be naturally introduced 
into the problem on hand. Assuming that the periastron distance
$r_{p}$ is sufficiently large, there is a small parameter in the
problem, namely the ratio of the typical frequency associated with
the periastron passage, $\Omega_{p}\sim \sqrt {GM\over r_{p}^{3}}$ to the
typical frequency associated with the planet, 
$\Omega_{*}=\sqrt{GM_{*}\over R_{*}^{3}}$: $\Omega_{p} \ll \Omega_{*}$
\footnote{Let us recall that $R_{*}$ and $M_{*}$ are the radius and
the mass of the planet, respectively, and $M$ is the mass of the 
object exerting tides on the planet.}. 
We would expect that the angular momentum transfer
between the orbit and the modes leads to a so called state of 
pseudo synchronisation where $\Omega \sim\Omega_{p}$, and therefore
the angular velocity of the planet can be, typically, much smaller than
the characteristic frequency $\Omega_{*}$. In such a situation all modes
of pulsation present in a convective planet can be divided in two distinct 
groups: high frequency modes with eigen frequencies of the order of 
or larger than $\Omega_{*}$, and 
low frequency modes  with eigen frequencies of the
order of $\Omega \sim \Omega_{p}$. 
For the high frequency modes, the  effects of rotation on the
structure of eigen functions and eigen frequencies
can be taken into account as a perturbation (see eg. IP) and references therein). Calculation 
of the eigen spectrum of the low frequency modes is much more
  difficult (Papaloizou $\&$ Pringle 1981, hereafter PP). 
However, a  convenient approximation can also be made in this case which 
allows one  to reduce the general equation $(\ref{eq p15})$ to a much
simpler equation for the low frequency modes only (PP, PI).

\subsection{Eigen problem for low frequency modes}

In order to reduce equation $(\ref{eq p15})$ to  an expression appropriate for low
frequency modes it is convenient to use the potential $\tilde W_{m}$ 
(see equation $(\ref{eq p4})$ for definition) as a dynamical variable
instead of the displacement vector $\tilde {\bmth {\xi}}_{m}$. 
We express the components 
of $\tilde {\bmth{\xi}}_{m}$ in terms of $\tilde W_{m}$ using  the
Fourier transformed variant of equation $(\ref{eq p3})$  having set 
${\bmth {f}}_{\nu}=0$ there. For a normal mode labelled by $k,$  the  components of the
displacement vector  are related 
to the potential through:
\begin{equation}
\xi_{\varpi, k}={1\over \sigma_{k} d_{k}}\left(-\sigma_{k} 
{\partial W_{k}\over \partial \varpi}+{2m\Omega\over \varpi}W_{k}\right ),
\label{eq p38}
\end{equation}
\begin{equation}
\xi_{\phi, k}={i\over \sigma_{k} d_{k}}\left(-{\sigma_{k} m\over \varpi} 
W_{k}+ 2\Omega {\partial W_{k}\over \partial \varpi}\right),
\label{eq p39}
\end{equation}
\begin{equation}
\xi_{z,k}={1\over \sigma_{k}^{2}}{\partial W_{k}\over \partial z},
\label{eq p40}
\end{equation}
where $d_k=4\Omega^{2}-\sigma_{k}^{2}$. To simplify notation
we   shall not use tildes  and subscripts denoting  $m$  for
quantities associated with eigen modes.

Now we substitute equations  $(\ref{eq p38})$, $(\ref{eq p39})$
and $(\ref{eq p40})$
into right hand side of the continuity equation  
$(\ref{eq p6})$. The density perturbation on the right hand side 
is expressed in terms of $W_{k}$ and perturbed gravitational potential 
$\Psi_{k}^{int}$ with help of equation $(\ref{eq p4})$ where we set
$\Psi^{ext}=0$. We   then obtain  
\begin{equation} 
\sigma_{k}^{2}  {\bmth{A}} W_{k} -\sigma_{k} {\bmth{B}}
W_{k} -{\bmth{C}} W_{k} =
\sigma_{k}^{2}d_{k}{\rho \over c_{s}^{2}}
(\Psi_{k}^{int}- W_{k}),  
\label{eq p41}
\end{equation}
and
\begin{equation} 
{\bmth {A}}=-{1\over \varpi}{\partial \over \partial \varpi}\left (\varpi \rho
{\partial\over \partial \varpi}\right )-{\partial \over \partial z}\left (\rho
{\partial \over \partial z}\right )+
{m^{2} \rho \over \varpi^{2}},  
\label{eq p42}
\end{equation}
\begin{equation} 
{\bmth {B}} =-{2m\Omega \over \varpi}{\partial \rho \over \partial \varpi}, 
\quad {\bmth {C}}=-4\Omega^{2}{\partial \over \partial z}\left(\rho {\partial
\over \partial z}\right).  
\label{eq p43}
\end{equation}
Note that the operator $\bmth {A}$ can be represented in an invariant form
\begin{equation} 
\bmth {A}= -\nabla^{\alpha}(\rho \nabla_{\alpha}).
\label{eq p43a}
\end{equation} 
As follows from equation $(\ref {eq p43a})$, $\bmth {A}$ is just a modified Laplace 
operator.

The operators ${\bmth {A }}$, ${\bmth {B}}$ and ${\bmth {C}}$ are 
analogous to the operators ${\bmth {\cal B}}$ and ${\bmth {\cal C}}$ 
introduced above.
They are self-adjoint with respect to the inner product
$(\ref{eq p2})$.
Also when $W$ is
not a constant
${\bmth {A}}$ and ${\bmth {B}}$ are positive 
definite and  ${\bmth {C}}$ is non negative definite. 

Note that an analogous equation used by PI has a factor 
$\sigma^{2}(4\Omega^{2}-\sigma^{2})$ instead of $\sigma_{k}^{2}d_{k}$
on the right hand side. Accordingly, the potential $W^{PI}_{k}$ used
by PI is different from the potential $W_{k}$ used in this Paper:
$W_{k}^{PI}=({\sigma \over \sigma_{k}})^{2}({d\over d_{k}})W_{k}$.

Equation for the perturbed gravitational potential $\Psi_{k}^{int}$ is 
obtained from equations $(\ref{eq p4})$ and $(\ref{eq p7})$
\begin{equation} 
\Delta \Psi_{k}^{int}+{4\pi G\rho\over c_{s}^{2}}\Psi_{k}^{int}=
{4\pi G\rho\over c_{s}^{2}}W_{k}
\label{eq p44}
\end{equation}

Equations $(\ref{eq p41})$ and $(\ref{eq p44})$ form a complete
set. They are equivalent to equation $(\ref{eq p18})$ and 
can be used instead of it to 
find the solutions for the eigen modes. 

Assuming that $\sigma_{k} \sim \Omega $ it is easy to see that the terms
on the right hand side of equation $(\ref{eq p41})$ are much smaller
than the terms on the left hand side. In the leading approximation in
the small parameter $(\Omega /\Omega_{*})^{2}$ they can be neglected 
and we obtain
\begin{equation} 
\sigma_{k}^{2}  {\bmth{A}} W_{k} -\sigma {\bmth{B}}
W_{k} -{\bmth{C}} W_{k} =0.
\label{eq p45}
\end{equation}
Equation $(\ref{eq p45})$ can be used for calculation of the low 
frequency eigen functions and eigen frequencies. Then a formalism exactly equivalent to that given by PI is obtained.

\subsubsection{Equation $(\ref{eq p45})$ as a self adjoint problem}
 
Equation $(\ref{eq p45})$ can be brought into the standard self adjoint
form analogous to equation $(\ref{eq p18})$ (PI). For that let us consider
two dimensional vector $\vec Y_{k}$ with components
\begin{equation}
Y^1_{k}  =  \sigma {\bmth {A}}^{1/2} W_{k}, \quad
Y^2_{k}={\bmth {C}}^{1/2} W_{k}.
\label{eq p46}
\end{equation}
Now it follows from  $(\ref{eq p45})$ that it can be written in the
form
\begin{equation}
\sigma_{k} \vec Y_{k} ={\bmth {H}} \vec Y_{k},  
\label{eq p47} 
\end{equation}
where
\begin{equation}
{\bmth {H}}=  
\left( \begin{array}{cc} {\bmth {A}}^{-1/2}{\bmth {B}}
{\bmth {A}}^{-1/2} & 
{\bmth {A}}^{-1/2} {\bmth {C}}^{1/2} \\
{\bmth {C}}^{1/2} {\bmth {A}}^{-1/2} & 0 \end{array}\right). 
\label{eq p48}
\end{equation}
Since the off diagonal terms in the matrix  $(\ref{eq p48})$ are
adjoint of each other and the diagonal terms are self adjoint it is
clear that the operator ${\bmth {H}}$ has the required self adjoint
form.

The inner product induced by the operator ${\bmth {H}}$ is  
\begin{equation}
<<\vec Y_{k}| \vec Y_{l}>>=\sigma_{k}\sigma_{l}(W_{k}| {\bmth {A}}
W_{l})+(W_{k}|{\bmth {C}} W_{l}).  
\label{eq p49}
\end{equation} 
It is obvious that different solutions of equation  $(\ref{eq p47})$ are
orthogonal with respect to this product.
 
From equations $(\ref{eq p21a})$, $(\ref{eq p38}-\ref{eq p40})$  
and $(\ref{eq p49})$ it follows that
the norm $N_{k}$ corresponding to the displacement vector associated
with $W_{k}$ is related to the norm $n_{k}=\sigma_{k}^{2}(W_{k}| {\bmth {A}}
W_{k})+(W_{k}|{\bmth {C}} W_{k})$ induced by the inner product
$(\ref{eq p49})$ as
\begin{equation}
N_{k}=n_{k}/(\sigma_{k}^{2}d_{k}).
\label{eq p49a}
\end{equation}

The density perturbation determined by a solution of equation
$(\ref{eq p47})$ follows from equation $(\ref{eq p4})$
\begin{equation}
\rho^{'}={\rho \over c_{s}^{2}}(W_{k}-\Psi^{int}_{k}),
\label{eq p50}
\end{equation} 
where the gravitational potential $\Psi^{int}_{k}$ is found 
from equation $(\ref{eq p44})$ with $W_{k}$ as a source.

The coefficients $S_{k}$ determining the energy and angular momentum 
transfer (see equations $(\ref{eq p28})$ and $(\ref{eq p35})$) 
follow from equations $(\ref{eq p22})$ and $(\ref{eq p50})$.  
We have
\begin{equation}  
S_{k}=\int \varpi d\varpi dz {\rho \over c_{s}^{2}}
(W_{k}-\Psi_{k})\Psi^{ext},
\label{eq p51}
\end{equation} 
where we take into account that we can use real expressions for
the potentials $W_{k}$ and $\Psi^{ext}$.

The expression $(\ref{eq p51})$ can be brought in a simpler form. For
that, let us formally solve equation $(\ref{eq p44})$
\begin{equation} 
\Psi_{k}^{int}=4\pi G \Delta_{*}^{-1} {\rho \over c_{s}^{2}}W_{k},
\label{eq p52}
\end{equation} 
where $\Delta_{*}^{-1}$ is the inverse of the operator 
$\Delta_{*}=\Delta+{4\pi G\rho\over c_{s}^{2}}$. 
It is easy to see that $\Delta_{*}^{-1}$ 
is self adjoint. We may, therefore, write the last term 
on the right hand side of $(\ref{eq p51})$ as 
\begin{equation} 
\int \varpi d\varpi dz {\rho \over c_{s}^{2}}\Psi_{k}\Psi^{ext}=
4\pi G\int \varpi d\varpi dz {\rho \over c_{s}^{2}}
(\Delta_{*}^{-1} {\rho \over c_{s}^{2}}W_{k})\Psi^{ext}=
-\int \varpi d\varpi dz {\rho \over c_{s}^{2}}W_{k}\Psi^{1},
\label{eq p53}
\end{equation}
where the potential $\Psi^{1}$ is determined from equation
analogous to equation $(\ref{eq p44})$ but with negative of
the potential $\Psi^{ext}$ as a source
\begin{equation} 
\Delta \Psi^{1}+{4\pi G\rho\over c_{s}^{2}}\Psi^{1}=
-{4\pi G\rho\over c_{s}^{2}}\Psi^{ext}.
\label{eq p54}
\end{equation}
The 'polarisation' potential $\Psi^{1}$ has a simple physical 
meaning. If the external potential $\Psi^{ext}$ is assumed to be 
static, the potential $\Psi^{1}$ is induced in the star by the
corresponding perturbation of density. 

Finally we obtain
\begin{equation}  
S_{k}=\int \varpi d\varpi dz {\rho \over c_{s}^{2}}W_{k}(\Psi^{ext}+\Psi^{1}).
\label{eq p55}
\end{equation}

\section{Explicit expressions for the energy and angular momentum transfer}

In this Section we obtain explicit expressions 
for the transfer of energy and angular momentum in terms of quantities
characterising the perturbing tidal field and quantities
characterising the planet. These  will be  applied directly
to the problem of evaluating the tidal response during a fly by
or as a consequence of multiple encounters during circularisation
from a highly eccentric orbit.  Since the expressions for the
energy transfer for the fundamental and $p$ modes are well known (see
PT, PI, IP and references therein) we consider only the low frequency
modes  in this section.

\subsection{Fourier transform of the tidal potential}    

As we have mentioned before we consider highly eccentric orbits 
in this Paper, with eccentricity $e \sim 1$. In this case 
the energy and angular momentum transfer takes place mainly near the
periastron $r_{p}$ and we make the usual approximation of the orbit as a 
parabolic one with the same value of $r_{p}$ for calculation of the
perturbing potential $\tilde \Psi^{ext}_{m}$.

The Fourier's transform of the perturbing potential $\tilde \Psi^{ext}$ 
for the parabolic orbit in the quadrupole approximation
has been calculated by PT. We have 
\begin{equation}
\tilde \Psi^{ext}_{m}=\sqrt{{5\over 2\pi^{3}}{(2-|m|)!\over (2+|m|)!}}            
{B_{|m|}\over
(1+q)}I_{2,-m}(y) \Omega_{p}r^{2}P^{|m|}_{2}(\cos \theta ),  
\label{eq p56} 
\end{equation} 
where $B_{2}=\sqrt{{3\pi\over 10}}$, $B_{0}=-{1\over 2}\sqrt{{\pi\over 5}}$.
$q$ is the ratio of $M_{*}$
to the mass of the star exerting the tides, $M$: $q=M_{*}/M$. In the
application of our formalism to the problem of circularisation of the
extra solar planets $q \sim 10^{-3}$ is very small and will be
neglected later.
\begin{equation} 
\Omega_{p}=\sqrt{{GM(1+q)\over r^{3}_{p}}}
\label{eq p57} 
\end{equation}
is a typical frequency of
periastron passage. The
functions 
\begin{equation}
I_{2,-m}(y)=\int^{\infty}_{0}dx(1+x^{2})^{-2}\cos [\sqrt 2
y(x+x^{3}/3)+2m\arctan (x)]
\label{eq p58} 
\end{equation}
specify the dependence of $\tilde \Psi^{ext}_{m}$ on $\sigma$, they are
described by PT. 
y=$\bar \Omega (\bar \sigma +m)$,
where we use dimensionless frequencies 
\begin{equation}
\bar \sigma =\sigma/ \Omega, \quad \bar\Omega =\Omega/\Omega_{p}.
\label{eq p59} 
\end{equation} 
$P_{l}^{m}(x)$ is the associated Legendre function, 
and $(r,\theta)$ are the spherical coordinates.  

When $y \gg 1$ the quantities $I_{2,-m}(y)$ decay exponentially. This
is valid for the case of the high frequency fundamental and $p$ modes 
where $\bar \sigma_{k}\bar \Omega  \gg 1$
for the eigen frequencies. 
On the other hand, the low frequency
modes considered above have $\bar \sigma_{k} \sim 1$ and  
do not experience this suppression. 

\subsection{First passage problem}\label{fpp} 

The expressions for the energy and angular momentum transfer $(\ref{eq p33})$ 
and $(\ref{eq p35})$ are strictly valid only under assumption that the 
planet is unperturbed before the fly by, and we call the problem
formulated in such a way as a 'first passage' problem.
This assumption may not be correct when 
the orbit of the planet is elliptical and 
the viscous time $t_{\nu}=1/\sigma_{\nu}$ is larger than the orbital
period $P_{orb}$. In latter case the energy and angular momentum
exchange between the orbit and the modes depends on the state of the
pulsational modes before the periastron passage. However, the
expressions for the first passage problem provides a basis for a
treatment of that more complicated case, and firstly we would like to discuss
this problem. The generalisation to the case of a multi-passage problem 
is discussed in the next Section.

In our discussion below we assume that effects determined by 
perturbations of the structure
of the planet due to rotation are not significant for our purposes.
We use, accordingly,
spherical models of the planet in order to find the   form of
the density and the sound speed entering in equations 
$(\ref{eq p29})$, $(\ref{eq p35})$. In the same approximation, taking
into account that the perturbing potential is proportional to the
product of a function of the radius $r = \sqrt{\varpi^2 +z^2}$ 
and a spherical harmonic, we see
that the variables in equation $(\ref{eq p54})$ can be separated in 
spherical polar coordinates. Therefore we can express the sum $\Psi^{ext}+\Psi^{1}$
entering in equation $(\ref{eq p55})$ as 
\begin{equation}
\Psi^{ext}+\Psi^{1}=F(r)\Psi^{ext},
\label{eq p59a} 
\end{equation} 
where the function  $F(r)$ can be obtained from
equation  $(\ref{eq p54})$. 
  
In order to present the expressions $(\ref{eq p29})$, $(\ref{eq p35})$  
and $(\ref{eq p37})$ in a simpler form we  express the
quantities $E_{c}$, $E_{I}$  and $L_{c}$ in terms of 
natural units of energy $E_{*}=GM_{*}^{2}/R_{*}$ and
angular momentum $L_{*}=M_{*}\sqrt{(GM_{*}R_{*})}$. As above we also  adopt
dimensionless units for the spatial coordinates, density and the sound 
speed expressing them in units of $R_{*}$, 
the averaged density 
${\bar \rho}={3M_{*}\over 4\pi R_{*}^{3}}$, and $\sqrt {GM_{*}\over
R_{*}}$, respectively. 

Instead of the separation distance $r_{p}$ it is convenient to
introduce the parameter (PT) 
\begin{equation}
\eta ={\Omega_{*}/ \Omega_{p}}=
\sqrt{{1\over (1+q)}{M_{*}\over M}{r_{p}^{3}\over R_{*}^{3}}}.
\label{eq p61} 
\end{equation}   

We then substitute equations  $(\ref{eq p55})$ and  $(\ref{eq p59a})$
into  $(\ref{eq p28})$ and  $(\ref{eq p35})$.
We obtain  
\begin{equation} 
E_{m}={C_{m}\over (1+q)^{2}}\bar {\Omega}^{4}\sum_{k} \lbrace 
\bar \sigma_{k}^{4}(4-\bar \sigma_{k}^{2})Q_{k}^{2}
I^{2}_{2,-m}(y_k) \rbrace {E_{*}\over \eta^{6}},  
\label{eq p62}
\end{equation}
\begin{equation}
L_{2}={2C_{2}\over (1+q)^{2}}\bar {\Omega}^{3}\sum_{k} \lbrace 
\bar \sigma_{k}^{3}(4-\bar \sigma_{k}^{2})Q_{k}^{2}
I^{2}_{2,-m}(y_k) \rbrace {L_{*}\over \eta^{5}},  \label{eq p63}
\end{equation} 
where we separate contributions to $E_{c}$ and $L_{c}$ corresponding
to different values of $m$ such as $E_{c}=\sum_{m=0,2}E_{m}$ and take
into account that $L_{0}=0$ and therefore $L_{c}=L_{2}$. All
quantities in equations  $(\ref{eq p62})$, $(\ref{eq p63})$ are
assumed to be in the dimensionless form, $C_{2}={3\over 16},$ 
$C_{0}=3/4$ and
\begin{equation}
y_k =\bar \Omega (\bar \sigma_{(k)} +2). 
\label{eq p90}
\end{equation}

The overlap integrals 
$Q_{k}$ are analogous to the same quantities introduced by PT
in their analysis of dynamic tides in
a non rotating star. They describe spatial coupling of 
the perturbing potential and of the eigen modes, and have the form:
\begin{equation}
Q_{k}=({\rho\over c_{s}^{2}}F(r)r^{2}P^{m}_{2}|W_{k})/\sqrt{\bar n_{k}}, 
\label{eq p64}
\end{equation}
where $\bar n_{k}=n_{k}/\Omega^{2}$.

The expression  $(\ref{eq p62})$ and $(\ref{eq p63})$ have already
been obtained by PI using a different method. They can also be presented
in another equivalent form as
\begin{equation}
E_{m}={\epsilon_{m}(\bar \Omega)\over (1+q)^{2}\eta^{6}}E_{*}, 
\quad
L_{2}={\lambda(\bar \Omega)\over (1+q)^{2}\eta^{5}}L_{*},
\label{eq p65}
\end{equation}
where the dependence of the functions $\epsilon_{m}$ and $\lambda$
on $\bar \Omega$ follows from $(\ref{eq p62})$ and $(\ref{eq p63}).$
Accordingly, the energy transfer in the inertial frame follows from
$(\ref{eq p37}),$  $(\ref{eq p65})$ and the use of 
\begin{equation}
\epsilon_{m}=C_m\bar\Omega^4\sum_{(k)}  
\bar \sigma_{(k)}^{4}(4-\bar \sigma_{(k)}^{2})Q_{(k)}^{2}
I^{2}_{2,-m}(y_k),\quad
\label{eq p891}\end{equation}
where all quantities in the brackets belong to a particular 
value of $m =0,2,$
\begin{equation}
\lambda = 2C_2\bar \Omega^3\sum_{(k)}  
\bar \sigma_{(k)}^{3}(4-\bar \sigma_{(k)}^{2})Q_{(k)}^{2}
I^{2}_{2,-2}(y_k), 
\quad\label{eq p892}\end{equation}
where all quantities in the brackets are evaluated for $m =2$
and 
\begin{equation}
\epsilon_{I} =\epsilon_{2}+\bar \Omega \lambda, 
\label{eq p89}
\end{equation}
as
\begin{equation}
E_{I}={(\epsilon_0+\epsilon_{2}+\bar \Omega \lambda)\over 
(1+q)^{2}\eta^{6}}E_{*} \  =  \ 
{(\epsilon_0+\epsilon_{I})\over 
(1+q)^{2}\eta^{6}}E_{*}.
\label{eq p66}
\end{equation} 

Usually, the moment of inertia of the planet is much smaller than
the moment of inertia of the orbit. In this case the planet tends to
rotate with 'equilibrium' value of the angular velocity, $\Omega_{eq}$
determined by condition that there is no exchange of angular momentum
between the planet and the orbit at this rotation rate,
\begin{equation}
\lambda (\Omega_{eq})=0.
\label{eq p67}
\end{equation} 
This state of rotation is often referred to as a state of 
'pseudo-synchronisation'. For a given model of planet, equation 
$({\ref {eq p67}})$ can be solved numerically to find $\Omega_{eq}$.
The expression for the energy transfer at the state of
'pseudo-synchronisation', $E_{ps}$, follows from equation 
$({\ref {eq p66}})$
\begin{equation}
E_{ps}={\epsilon_{*}\over \eta^{6}}E_{*},
\label{eq p68}
\end{equation} 
where
$\epsilon_{*}=\Sigma_m \epsilon_{m}(\Omega_{eq})$
depends only on the structure of a planet. For simplicity we set $1+q = 1$
hereafter. The expressions for a comparable mass ratio $q$ can be
easily obtained from our equations by an appropriate re-scaling of
variables we use.  

An estimate of $\epsilon_{*}$ can be obtained from the following 
semi-analytical arguments.
As we will see below only two 'global' modes with $m=2$ mainly 
determine the energy transfer when $\Omega \sim \Omega_{eq}$
and we can take into account only these two modes in the summation
 $(\ref {eq p67})$ and $(\ref {eq p68})$.
One of these modes has a positive eigen 
frequency with $\bar \sigma_{1}\approx 0.6$ and
the overlap integral $Q_{1}\approx 0.16$, and another mode has
a negative eigen frequency with $\bar \sigma_{2}\sim -1.1$. The
equilibrium value of the angular velocity is always close to
the frequency of periastron passage, with $\bar \Omega_{eq}\approx 1.55$.
This information is sufficient to estimate $\epsilon_{*}$. Indeed, we 
can express the second overlap integral from equation $(\ref {eq p67})$
in terms of $Q_{1}$, the eigen frequencies and the quantities
$I_{2,-2}(y_{1,2})$, where $y_{1,2}=\bar \Omega_{eq}
(2+\sigma_{1,2})$. We get $Q_{2}\sim 0.033$.
Taking into account that $I_{2,-2}(y_{1})\approx 0.32$ and 
$I_{2,-2}(y_{2})\approx 0.74$, we get $\epsilon_{*}\sim 4\cdot
10^{-3}$. This value is in a good agreement with numerical results
discussed below, see discussion and equation $(\ref {eq p94})$ 
in Section 6. Note that for $n=1.5$ polytrope 
$\epsilon_{*}\approx 6.5 \cdot 10^{-3}$ (PI).
 
\subsection{Multi-passage problem}

In the circularisation problem expressions $(\ref {eq p28})$,
$(\ref {eq p35})$, $(\ref {eq p37})$, $(\ref {eq p62})$, 
$(\ref {eq p63})$, $(\ref {eq p65})$, $(\ref {eq p66})$ 
and $(\ref {eq p68})$ are strictly
valid only when the energy stored in the modes can be dissipated   between
two successive periastron passages. In the opposite limit, when the
dissipation time $t_{\nu} \gg P_{orb}$, where  $P_{orb}$
is the orbital
period, the planet approaches periastron in a perturbed state. The
additional linear perturbation induced during a particular periastron 
passage must be added to the perturbation already present in the star. Since
the mode energy is quadratic in the perturbed quantities, it can 
either decrease or increase after the periastron passage, depending on
phase of the perturbations (Kochanek, 1992, Mardling 1995a, 1995b). 
This issue has been studied analytically by IP. It was found that when
a change of the orbital period between two successive periastron
passages is sufficiently large, the energy transfer proceeds in a 
stochastic way. In this case the phase of perturbation in the planet
approaching periastron is not correlated with the phase of the 
perturbation induced by tides near  periastron, and the energy
transfer during the periastron passage is essentially a random quantity,
with dispersion of the order of the expressions for the energy
transfer calculated above. In this situation, a build-up
of mode energy due to stochastic instability is possible. A
semi-analytical condition for this instability to happen has been
obtained in IP. Here we  
generalise their formalism to take account of the
specific features associated with  inertial waves.

Let us assume that dissipation is negligible and treat the tidal 
influence as a sequence of impulses separated by periods of time
equal to the orbital period. In this case,  any time between two
impulses the planet  undergoes free oscillation, and
the Lagrangian displacement vector can be represented by 
an expression analogous to $(\ref {eq p26})$
\begin{equation}
{\bmth {\xi}}=2\pi i\sum_{m,k} ({\sigma_{k} S_{k}\over N_{k}}
A_{k}e^{i\psi_{k}}
e^{-i\sigma_{k}t}e^{im\phi}{\bmth {\xi}}_{k}+c.c.),
\label{eq p69} 
\end{equation}
where the dimensionless real coefficients $A_{k} \ge 1$ characterise the
amplitudes of perturbations corresponding to different modes 
and  $\psi_{k}$ are the corresponding phases. 
Note that the origin of time is chosen to coincide with the moment of
periastron passage. The expressions for the canonical energy and
angular momentum follow from $(\ref {eq p27})$, $(\ref {eq p34})$ and
$(\ref {eq p69})$
\begin{equation}
E_{c}=8\pi^{3}\sum_{m,k}A_{k}^{2}{\sigma_{k}^{2}S_{k}S_{k}^{*}\over N_{k}}, \quad
L_{c}=16\pi^{3}\sum_{m=2,k}A_{k}^{2}\sigma_{k}{S_{k}S_{k}^{*}\over N_{k}}.
\label{eq p70} 
\end{equation}
When $A_{k}=1$ these expressions obviously coincide with the
expressions $(\ref {eq p28})$ and $(\ref {eq p35})$  for the single
passage problem. For  inertial waves equations $(\ref {eq p70})$
can be brought to the form analogous to equations $(\ref {eq p65} -\ref{eq p892})$
\begin{equation}  
E_{m}=\left(\sum_{k}\epsilon_{m,k}A_{k}^{2}\right){ E_{*}\over (1+q)^{2}\eta^{6}}, 
\quad
L_{2}=\left(\sum_{k} \lambda_{k}A_{k}^{2}\right){ L_{*}\over (1+q)^{2}\eta^{5}}, 
\label{eq p71} 
\end{equation}
where we now show separately the contributions of terms with different
$m$ to the canonical energy. $\epsilon_{m,k}$ and  $\lambda_{k}$ are 
the terms in the summation series for  $\epsilon_{m}$ and  
$\lambda$, corresponding to a particular value of $k$, see equations
$(\ref{eq p65})$, $(\ref{eq p891})$ and $(\ref{eq p892})$. 
As we discus   below in the tidal problem we can neglect the
contribution of $m=0$ terms. In this case, the mode energy in the 
  inertial frame can be represented as
\begin{equation} 
E_{I}=\left(\sum_{m=2, k}{(\epsilon_{2,k}+\bar \Omega \lambda )\over
\epsilon_{*}}
A_{k}^{2}\right)E_{ps},
\label{eq p72} 
\end{equation}
where $E_{ps}$ is given by $(\ref {eq p68})$

After the periastron passage, the expression  
$(\ref {eq p26})$ must be added to $(\ref {eq p69})$, and the
resulting expression can brought again to the form $(\ref {eq p69})$
but with new values of the amplitudes and phases, $A^{n}_{k}$ and
$\psi^{n}_{k}$. Thus, the action of tidal forcing defines an iterative 
map $(A^{n-1}_{k}, \psi^{n-1}_{k}) \rightarrow (A^{n}_{k},
\psi^{n}_{k})$. It is convenient to introduce new variables 
$(x^{n-1}_{k},x^{n}_{k})=(A^{n-1}_k\cos \psi_k^{n-1}, A_k^{n}\cos \psi_k^{n})$ 
and
 $(y^{n-1}_{k}, y^{n}_{k})=
( A_k^{n-1}\sin \psi_k^{n-1}, A_k^{n}\sin \psi_k^{n} )$. In terms of these variables
this map has especially simple form (IP)
\begin{equation}
\left( \begin{array}{cc} x^{n}_{k} \\ y^{n}_{k} \end{array} \right)
={\bmth {R}}(-\phi_{k})
\left( \begin{array}{cc} x^{n-1}_{k}+1 \\ y^{n-1}_{k} \end{array} \right),
\label{eq p73}
\end{equation}  
where $\phi_{k}=\sigma_{k}P_{orb}$, and ${\bmth {R}}(\phi)$ is
a two dimensional rotational matrix. Note that in 
equation $(\ref {eq p73})$ we take into account the change of
origin of time after the periastron passage: $t \rightarrow t+P_{orb}$.

When $P_{orb}$ is a constant, it is easy to see from equation 
$(\ref {eq p73})$ that the amplitudes $A^{n}_{k}$ do not grow with 
the number of iterations. However, the orbital period depends on 
the orbital  energy, $E_{orb}$, as
\begin{equation}
P_{orb}={\pi GM^{5/2}q^{3/2}\over \sqrt {2|E_{orb}|} |E_{orb}|(1+q)^{1/2}}, 
\label{eq p74} 
\end{equation}  
and the orbital energy is proportional to the mode energy $E_{I}$ due
to the law of energy conservation. Since 
$E_{I} = \left( \sum_{k,2}{(\epsilon_{2,k}+\bar \Omega \lambda )\over
\epsilon_{*}}
(A^{n-1}_{k})^{2}\right)E_{ps}$ is proportional to
the squares of the modes amplitudes, $A^{n-1}_{k}$, the map is non-linear.

In order to simplify the problem, let us consider a situation when
the mode energy is much smaller than the orbital energy such that the orbital 
  energy and angular momentum 
remain in the neighbourhood of some particular  fixed values. In this
case we can set $P_{orb}=P_{orb}(0)+\Delta P_{orb}$, where
$P_{orb}(0)$ is a constant, and 
\begin{equation} 
\Delta P_{orb}= -{3\over 2}{P_{orb}\over |E_{orb}|}{E_{I}\over
 M_{*}}=-6\pi \sqrt {a^{5}\over (GM)^{3}}{E_{I}\over M_{*}},
\label{eq p75} 
\end{equation} 
where $E_{orb}$ and $a$ are the orbital energy and semi-major
axis corresponding to the period $P_{orb}(0)$. Now the angles
$\phi_{k}$ can be expressed in terms of $E_{I}$ as
\begin{equation} 
\phi_{k}=\phi_{k}(0)-\bar \sigma_{k}\alpha \tilde E_{I},
\label{eq p76} 
\end{equation} 
where $\tilde E_{I}=E_{I}/E_{ps}$, 
$\phi_{k}(0) =\sigma_{k}P_{orb}(0)$ and
\begin{equation} 
\alpha=6\pi \epsilon_{*}\bar \Omega {M_{*}\over M}{1\over \eta^{6}}
\left({a^{5/2}\over r_{p}^{3/2}R_{*}}\right),
\label{eq p77} 
\end{equation}
where we use equations $(\ref {eq p59})$, 
$(\ref {eq p61})$, $(\ref {eq p68})$ and $(\ref {eq p75})$.
Equation  $(\ref {eq p77})$ is analogous to equation $(84)$ of IP. 
 
Equations $(\ref {eq p72})$, $(\ref {eq p73})$ and $(\ref {eq p77})$
fully determine the map for given values of $\phi_{k}(0)$ $mod$
$2\pi$ and $\alpha$.  When $\alpha$ are sufficiently large,
 stochastic instability sets in. Since the angles $\phi_{k}(0)$ 
are determined by initial conditions, the results should be averaged
over these angles. 

\begin{figure}
\begin{center}
\vspace{8cm}\includegraphics{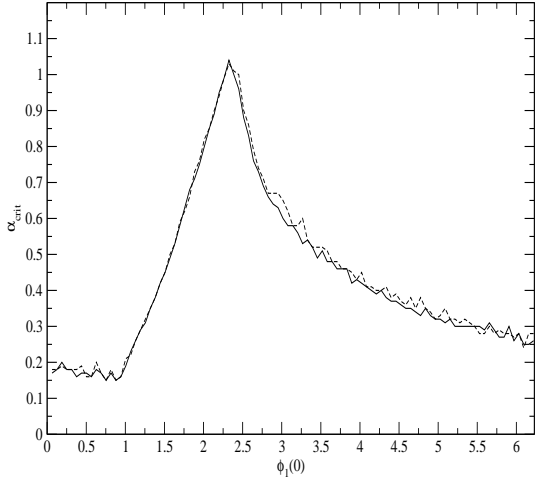}
\end{center}
\vspace{-1cm}
\caption{The dependence of $\alpha_{crit}$ on  $\phi_{1}(0)$. The solid
and dashed lines correspond to the maximal number of iterations
$N_{max}=10^{5}$ and $N_{max}=10^{4}$, respectively.}
\label{Fig.1}
\end{figure}

\begin{figure}
\begin{center}
\vspace{8cm}\includegraphics{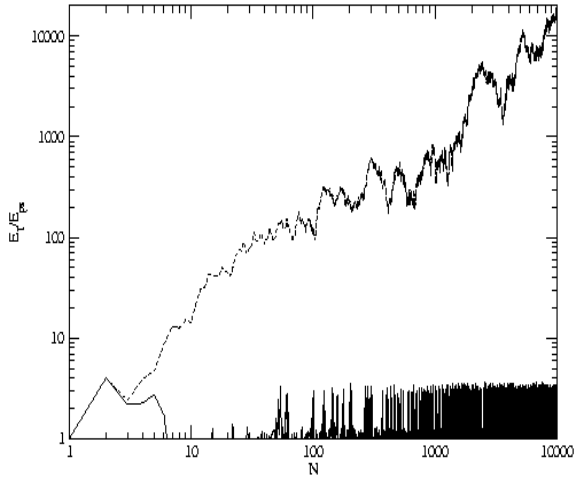}
\end{center}
\caption{$\tilde E_{I}$ as a function of the number of iterations $N$.
The case $\phi_{1}(0)=2.32$ corresponding to the maximal value of
$\alpha_{crit}=1.04 $ is shown. The solid curve corresponds to
$\alpha=0.9$ and the dashed curve corresponds to $\alpha=1.1$.}
\label{Fig.2}
\end{figure}

This map has been investigated numerically by IP for the case of only
one high frequency mode. It has been found that the instability sets in when the 
average value of $\alpha$ is larger than $\alpha_{crit}\approx 0.67$.
In our case the situation is slightly more complicated since 
there are at least two modes which play a role in the
energy transfer. Also, we should specify the evolution of the angular 
velocity. In order to investigate this question we  
consider numerically the simplest possible case of 
constant equilibrium rotation of a planet,
when $\bar \Omega =\bar \Omega_{eq}$. For definiteness, 
we use parameters described in
the end of the previous section, namely $\bar \Omega_{eq}=1.55$ 
and consider only  $m=2$ modes with 
$\bar \sigma_{1}=0.6$ and $\bar \sigma_{2}=-1.1$. In this case we
have $\epsilon_{2}\approx 0.35 \epsilon_{*}$  in the first case and 
$\epsilon_{2}\approx 0.65 \epsilon_{*}$ in the second. We iterate numerically the 
map for a large number $N$ of iterations for a given value of
$\phi_{1}(0)$ with $\phi_{2}(0)={\bar \sigma_{2}
\over \bar \sigma_{1}}\phi_{1}(0)$, and for a given value of
$\alpha$. The stochastic instability 
is assumed to set in when the energy $\tilde E_{I}$ exceeds the 
maximal number of iterations, $N_{max}$ during the iterations.
This happens when $\alpha$ exceeds some critical value 
$\alpha_{crit}(\phi_{1}(0))$. 
The results of our computations are shown in Figure 1. One can see
from this Figure that there is a strong dependence of $\alpha_{crit}$ 
on  $\phi_{1}(0)$. Nevertheless we use the averaged value of
$\alpha_{crit}$, $\bar \alpha_{crit}={1\over
2\pi}\int^{2\pi}_{0}\alpha_{crit}\approx 0.43$ since our results are
rather insensitive to the exact values of $\alpha_{crit}$. It is
interesting to note that the transition from a quasi-periodic
behaviour of $\tilde E_{I}$ to  stochastic behaviour is very
sharp. In Figure 2 we show the evolution of $\tilde E_{I}$ for 
$\phi_{1}(0)=2.32$ and two values of $\alpha$: $\alpha=0.9 <
\alpha_{crit}=1.04$ and $\alpha=1.1 > \alpha_{crit}=1.04$. 
One can see from this Figure that 
a small change of $\alpha$ is sufficient for  
stochastic instability to set in.

From the condition $\alpha > \alpha_{crit}$ we find that the
stochastic instability sets in only when the semi-major axis is
sufficiently large,
\begin{equation}
a > a_{st}=({\alpha_{crit} \over 6\pi \bar
\Omega_{ps}\epsilon_{*}})^{2/5}\left({M\over M_{*}}\right)^{3/5}\eta^{14/5}R_{*}.
\label{eq p78}
\end{equation} 
Using the values of all quantities in $(\ref{eq p78})$
typical for the problem
of circularisation of extra solar planets, we have
\begin{equation} 
a_{st}\approx 30 \left({M_{J}\over M_{*}}\right)^{3/5}\left({M\over M_{\odot}}\right)^{3/5}
({R_{*}\over R_{J}})\eta_{10}^{14/5}au,
\label{eq p79}
\end{equation}
where $\eta_{10}=\eta/10$, $M_{J}\approx 2\cdot 10^{30}g$ and 
$M_{\odot}\approx 2\cdot 10^{33}g$ are the masses of Jupiter and of the
Sun, respectively, and $R_{J}\approx 7\cdot 10^{9}cm$ is the radius of
Jupiter. Remarkably, the typical value of $a_{st}\sim 30 au$ is
close to what has been obtained in the analysis of the stochastic 
instability associated with the fundamental modes (IP). This value 
is of the order of a typical 'initial' semi-major axis 
from which we would expect the process of circularisation to begin for $\eta =10.$
This corresponds to a final circularised period of $3.3$~d for a Jupiter mass planet around a
solar type star. However if the final period is reduced to $\sim 1,2$~d, corresponding
to the shortest period exoplanets,
the initial semi-major axis is reduced to between $1$ and $2 au.$
If the initial semi-major axis is smaller, the stochastic growth of
the mode energy is not possible. However, in this case, we would 
expect a stronger dissipation  of the mode energy due  non-linear
mode-mode interactions (e.g. Kumar $\&$ Goodman 1996). This possibility
needs a further investigation.

\section{Planet models and their normal mode spectra} 

In this Section we  construct realistic
models of  planets  at different evolutionary stages
for which we obtain the oscillation spectra numerically.
We discuss our numerical method, use it to  solve equation
$(\ref {eq p47})$, and discuss the results which are used in subsequent
sections to evaluate the tidal responses of different planet models
undergoing a fly by or circularisation from a highly eccentric orbit.

We find it convenient to  adopt a system of 
dimensionless units for the spatial coordinates, density and the sound
speed expressing them in units of $R_{*}$,
the averaged density
${\bar \rho}={3M_{*}\over 4\pi R_{*}^{3}}$, and $\sqrt {GM_{*}\over
R_{*}}$, respectively.
We  also use dimensionless frequencies
\begin{equation}
\bar \sigma =\sigma/ \Omega, \quad \bar\Omega =\Omega/\Omega_{p}.
\label{eq p5900}
\end{equation}

\subsection{The planet models}
\begin{figure}
\begin{center}
\vspace{8cm}\includegraphics{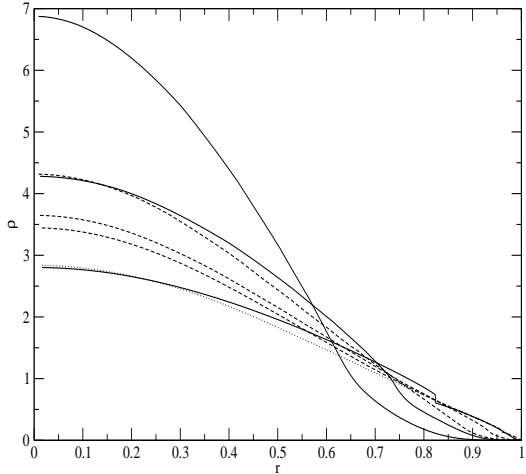}
\end{center}
\vspace{-1cm}
\caption{The  density $\rho $ scaled in terms of the mean density
as a function of the
dimensionless radius $r$. The three solid curves  are for models  with
$M_{*}=1M_{J}$  and radii $R_{*}=1R_{J}$, $1.4R_{J}$ and $2R_{J}$ respectively. 
The three dashed curved are for  models with 
$M_{*}=5M_{J}$ and radii $R_{*}=1.03R_{J}$, 
$1.4R_{J}$ and $2R_{J}$ respectively. The dotted curve corresponds to the model
with  $M_* = 1M_J$ and radius $R_* = 1R_J.$  and 'smoothed' density profile. 
Curves of the same type starting   from larger values  at $r=0$
 correspond
to  planet models with larger radii. 
Note that because of the use
of different scalings to dimensionless
variables  the numerical values  differ from  those of IP
by a factor ${ 4\pi \over 3}.$}
\label{Fig.3}
\end{figure}

\begin{figure}
\begin{center}
\vspace{8cm}\includegraphics{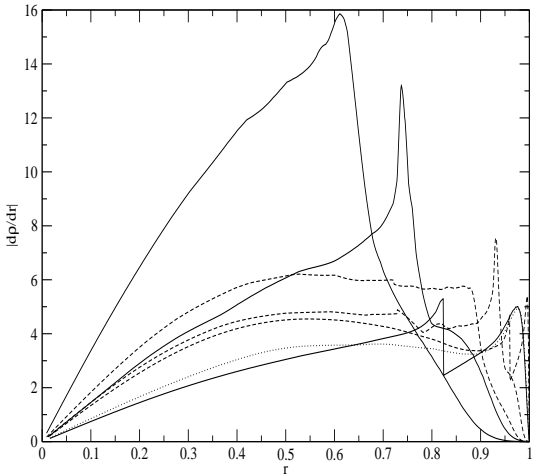}
\end{center}
\vspace{-1cm}
\caption{As in  Fig. \ref{Fig.3} but  the absolute value of the density gradient $|{d\rho
\over dr}|$ is shown. Curves of the same type  starting from larger values 
at $r = 0$  correspond to planet models with larger radii.}
\label{Fig.4}
\end{figure}

\begin{figure}
\begin{center}
\vspace{8cm}\includegraphics{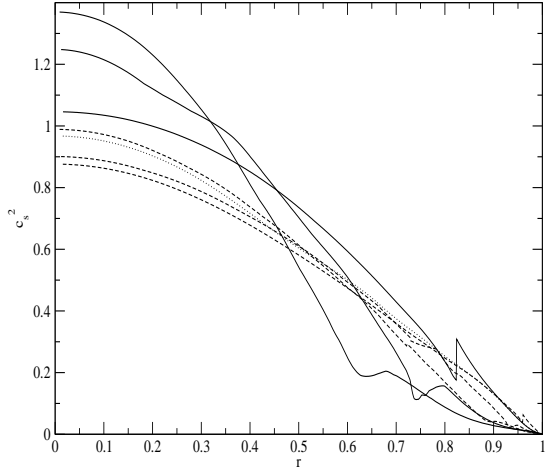}
\end{center}
\vspace{-1cm}
\caption{As for  Fig. \ref{Fig.3} and Fig. \ref{Fig.4} but the square of sound speed
$c_{s}^{2}$ is shown. Curves of the same type 
  starting from larger values  at $r = 0$
correspond to planet models with larger radii.}
\label{Fig.5}
\end{figure}

We assume that the planets are spherically symmetric and 
adopt the equation of state of Saumon, Chabrier $\&$ Van Horn (1995).
Although it is likely that giant planets form through
a nucleated core instability (eg. Bodenheimer $\&$ Pollack 1986),
for this first set of calculations, for reasons of simplicity
we shall assume that solid cores are absent.
In this context we note that the mass of any solid core in Jupiter
is highly uncertain, being sensitive among other things to the equation of state
adopted,  with recent estimates showing it to be  consistent
with zero (see Saumon \& Guillot 2004). 
To make this consistent with a nucleated core instability scenario
for formation, these authors suggest that a core could be eroded
and remixed during subsequent gas accretion of light elements.
  But it should be noted that 
the presence of a large solid core could significantly
affect the mode spectrum of the planet through the existence
of wave attractors (eg. Ogilvie \& Lin 2004) that could produce
effects on a long time scale. 
However, the presence of a small core  is unlikely to affect
phenomena of short duration such as the energy transfer that occurs
during a close passage or fly by.

We calculate
numerically the  form of $\rho$ and $c_{s}$  as functions of the 
radius $r.$  With these functions known, equation
$(\ref{eq p47})$ can be solved and the overlap
integrals $(\ref{eq p64})$ calculated. 
 As in IP we consider only two planet masses 
$M_{*}=1M_{J}$ and $5M_{J}$. 

For a given mass we 
calculate  planet models for six values of the radius.
These are  $R_{*}=1R_{J}$ 
($R_{*}=1.03R_{J}$) when $M_{*}=M_{J}$ ($M_{*}=5M_{J}$), and
$R_{*}=1.2R_{J}$, $1.4R_{J}$, $1.6R_{J}$, $1.8R_{J}$, $2R_{J}$ for both masses.
The different values of radius for a given mass
 correspond to different points
on an evolutionary track.
On such a track, the radius  decreases with time due to gravitational
contraction. To determine the age
corresponding to a given radius,
we use  the relationship between the radius and
luminosity, and hence the age of the planet calculated by Burrows
et al (1997) (see Fig. 8 of IP). 
Since a young hot planet of age $\sim 10^{6}yr$  
has a radius of around
$\sim 2R_{J},$ the range of radii we consider  corresponds to 
 planet ages between $\sim 10^{6}yr$  and
$\sim 5\cdot 10^{9}-10^{10}yr$. 

We illustrate the dependences of the density, the  magnitude 
of the density gradient
$|{d\rho \over dr}|$ and square of the sound speed for several
planet models in Figures \ref{Fig.3}-\ref{Fig.5}. As 
one can see from Figure \ref{Fig.3} for a particular planet,
the ratio of central density to  
 mean density increases with radius. This effect
is less prominent for the models with  $M_{*}=5M_{J}.$ 

\noindent Note a sharp drop of density at $r\approx 0.82$ when $M_* = 1M_J,$
and $R_* = 1R_J.$ 
This is due to the phase transition of
hydrogen from the metallic to molecular phase which is assumed to
be of the first kind. This phase transition also takes place 
when $M_{*}=5M_{J}$ and $R_{*}=1.03R_{J},$  but it is 
shifted to a larger $r\approx 0.96$ 
and  is not well resolved in this Figure.  In general, according to
Saumon, Chabrier $\&$ van Horn (1995), the phase transition is of the
first kind when $R_{*} < 1.27R_{J}$ ($R_{*} < 1.255R_{J}$) 
for $M_{*}=1M_{J}$ ($M_{*}=5M_{J}$). 
For larger radii the 
adiabats describing the interiors of the planets go above the 
critical point of Saumon, Chabrier $\&$ van Horn (1995)  equation of
state on $(P,T)$ phase plane. As we see below the effect of the phase
transition on the structure of the eigen spectrum  
is more prominent
for $M_{*}=1M_{J}$. 

\noindent To illustrate this it is of interest to compare this model
with a similar model without the phase transition and we  make such a 
model with 'smoothed out' transition between 
the metallic and molecular phases (dotted curves in Figures 3-5),  
for $M_{*}=1M_{J}$ and $R_{*}=1R_{J}.$
This model has been calculated using a prescription provided by
Saumon, Chabrier $\&$ van Horn 1995.

The structure of the density profiles is better seen in Figure \ref{Fig.4}
where the magnitudes of the density gradients are plotted.
 For the models with the
phase transition there is a discontinuity at the radius of the phase 
transition. For the models without the phase transition 
the modulus of the gradients increase with radius at small $r$, 
and they sharply decrease at large $r$. The transition point between
these two limiting cases corresponds to a 'melting' point where
the number density of the hydrogen molecules sharply decreases 
toward smaller values of $r$. This transition point shifts toward
smaller $r$ with increase of the planet radius.  As seen from Figure \ref{Fig.5}
 the sound speed is  changes non monotonically  near the transition point. 

In general, the models corresponding to the larger mass are more
similar to each other. In this case 
the radius of transition is shifted toward larger
$r$ where the planets have rather low density. 
The low density outer regions of the planet should not influence
 eigenfrequencies and eigenfunctions corresponding to global modes 
should they exist. Therefore,    
it is reasonable to
suppose the presence of the phase transitions and the 'melting'
zones in the  planet models with $M_{*}=5M_{J}$
is less significant for our problem than for the case of $M_{*}=1M_{J}$. 
As we see below this assumption is supported by our numerical results. 

\subsection{Numerical method}
For our numerical work we use a method similar to that 
proposed by PP.  This involves reducing the operator equations to discrete
form by use of a set  of  basis functions.

\subsubsection{Matrix representation of equations  $(\ref {eq p45})$
and  $(\ref {eq p47})$} 
For our numerical work it is convenient to work with a matrix
  representation of  equation
$(\ref {eq p45}).$   This is accomplished  by use of a basis of 'trial' 
functions  $v^{j}$, $j=1,..\infty$ which is assumed to be 
complete. We introduce a Fourier decomposition of an eigenfunction
$W_{(k)}$   in terms of the trial functions as
\begin{equation}
W_{(k)}=\sum_{j} \alpha^{(k)}_{j}v^{j},
\label{eq p80} 
\end{equation}
where the indices enumerating the eigen functions are enclosed in
 brackets from now on. Then we substitute equation $(\ref {eq p80})$
into equation $(\ref {eq p45})$ and taking the scalar product $(\ref {eq
p2})$ of the resulting expression with $v^{i}$ we get
\begin{equation}   
(\sigma_{(k)}^{2}A^{j}_{i}-\sigma_{(k)}B^{j}_{i}-C^{j}_{i})\alpha^{(k)}_{j}=0,
\label{eq p81} 
\end{equation}
where
\begin{equation}
A^{j}_{i}=(v^{i}|{\bmth {A}}v^{j}), \quad B^{j}_{i}=(v^{i}|{\bmth {B}}v^{j}),
\quad C^{j}_{i}=(v^{i}|{\bmth {C}}v^{j}), 
\label{eq p82}
\end{equation}
and the standard summation rule over the
repeating indices is implied
\footnote{ We stress that we do not sum over the indices enclosed in
the brackets that label an eigenfunction.}. 
We can introduce square roots and inverses of the above matrices in a
 similar manner  to that adopted for the corresponding operators. Let us
consider infinite dimensional vectors analogous to the vectors
$\vec Y_{(k)}$ (see $(\ref {eq p46})$), 
with components $Y^{(k)}_{M}$ where 
the index $M$ stands for the pair ($(1,i)$ and $(2,i)$). We have 
\begin{equation}   
Y^{(k)}_{1,i}=\sigma_{(k)}(A^{1/2})_{i}^{j}\alpha^{(k)}_{j}, \quad
Y^{(k)}_{2,i}=(C^{1/2})_{i}^{j}\alpha^{(k)}_{j}.
\label{eq p83}
\end{equation}
The eigen value problem $(\ref {eq p81})$ can be formulated for
the vectors $(\ref {eq p83})$ in a way equivalent to 
$(\ref {eq p47})$, 
\begin{equation}
\sigma_{(k)}Y^{(k)}_{N}=H_{N}^{M}Y^{(k)}_{M},
\label{eq p84}
\end{equation}
where the matrix $H_{N}^{M}$ has a structure equivalent to the 
operator ${\bmth {H}}$ (equation  $(\ref {eq p48})$) with the
operators being substituted by the corresponding matrices.

It is natural to use
the solutions of the eigenvalue problem
\begin{equation} 
{\bmth {A}}v^{j}=\lambda \rho v^{j}
\label{eq p85}
\end{equation} 
as the functions $v^{j}$. These functions are orthogonal with respect
to the inner product $(\ref {eq p2})$. 
As follows from equation $(\ref {eq p43a})$,
for the spherical planet models the variables in equation 
$(\ref {eq p85})$ are separable in  spherical  polar 
coordinates $(r, \theta )$, 
$v^{j}=v_{n,l}(r)P^{m}_{l}(\cos\theta)$, where $n=1,2,...$. $n-1$ 
is the  order of
the radial eigenfunction and the index $j$ stands for the pair of positive
integers $(n,l)$.
For the radial part we have
\begin{equation} 
-{1\over r^{2}}{d\over dr}\left(r^{2}\rho(r){d\over dr}v_{n,l}\right)+
{L^{2}\over r^{2}}\rho (r) v_{n,l} =\lambda_{n,l} \rho (r) v_{n,l},
\label{eq p86}
\end{equation} 
where $L=\sqrt{l(l+1)}$. 
It follows from  $(\ref {eq p86})$ that the
eigen values of ${\bmth {A}}$ are bound from below: 
$\lambda_{n,l} \ge L^{2}$ when $l\ne 0$. 

\subsubsection{Numerical approach}

In our numerical work we truncate the series $(\ref {eq p80})$ at some
sufficiently large value $j=j_{max}$. The trial functions must be 
arranged in such a way that when $j$ is sufficiently large the
corresponding trial functions oscillate with periods of spatial 
oscillation  that decrease with increasing $j$. 
In this case, one can argue that 
 errors   will be small because of a negligible contribution
of the oscillatory terms with  high $j$ to the 
various integrals   occurring in the formulation of our problem 
as discussed above. Of course,
this   idea must be tested, and we discuss   such tests later.

Since for our problem $m=0,2$ is
even  and the angular parts of the trial functions are such that 
$P^{m}_{l}(x)=(-1)^{l}P^{m}_{l}(-x),$  
the modes odd with respect to reflection in $x \rightarrow -x$ do not
couple with the tidal field  and  we can choose 
$l \ge m$ to be even. Thus $l=2l_{1}$ with $l_{1}$ being a positive integer
such that $l_{1} \ge m/2.$ 

\noindent  It is convenient to
consider an equal number of the radial functions $n(max)$ and 
angular functions $l_{1}(max).$  Thus we write
$n(max)=l_{1}(max)=N_{max}.$ In our numerical work, we use different
values of $N_{max}$ in order to study the dependence on the number of
 trial functions, with the largest $N_{max}=15$. 
Accordingly, we have  $j_{max}=N_{max}^{2} \le 225.$

After truncation of the   basis set used in $(\ref {eq p80})$ we have ordinary
finite dimensional matrices $A^{j}_{i}$, $B^{j}_{i}$ and
$C^{j}_{i}$. We  calculate the eigenvalues 
eigenvectors of these matrices. The eigenvalues are used to calculate
the inverses and square roots of the matrices entering in the matrix
$H_{N}^{M}$. We find the components of $H_{N}^{M}$ and solve  the
eigenvalue problem $(\ref {eq p84})$. The eigenvectors $Y^{(k)}_{N}$
are used to find the coefficients $\alpha^{(k)}_{j}$, and we use these 
coefficients to find the eigenfunctions $W_{(k)}$ with help of equation
$(\ref {eq p80})$. 

One can view the eigenvalues and eigenfunctions
as belonging to the oscillation spectrum of the planet.
For the case of a model without a solid core this
is ultimately likely to be discrete but everywhere dense (see eg. PP).
One should also consider
the possibility of the existence of  a continuous spectrum in some
cases. Then the numerically obtained eigenfunctions
may not show convergence to regular functions.
But note  that, independently of this issue,  we expect our numerical
method to be convergent when the eigenfunctions  we obtain are used
to provide a basis for  representing  smooth functions varying on a global
scale. The issue of the   possibility of ultimately
irregular eigenfunctions belonging to a continuous spectrum
does not affect this or  preclude the convergence of the  process
of calculating the angular momentum and energy
exchanges associated with a fly by.
On account of the  forcing by a global potential and short time duration of
a fly by, a basis representation capable of representing global
scales would be expected to be adequate on physical grounds.
Indeed for  our models  it turns out that only a few global eigenfunctions
are important for determining the energy  exchanged and these
can be represented with a basis set that needs only to be  capable
 of representing global functions.
This makes the issue qualitatively different to the one that
occurs in the context of tidal forcing occurring in a circular
orbit with fixed frequency for infinite time periods (eg.Ogilvie \& Lin
2004).

It is obvious that the dimension of the matrix $H_{N}^{M}$ is 
$2j_{max}$. Thus, we have $2j_{max}$ eigen values $\sigma_{(k)}$
and eigen vectors
\footnote{Note that some of $\sigma_{(k)}$ are
degenerate.}. The number of terms in   the series occurring in  
$(\ref {eq p62})$, $(\ref {eq p63})$ and  $(\ref {eq p71})$ 
determining the tidal response, is proportional to $j_{max}$. 
It is important to check whether the process adopted
and the  corresponding series
converges and we discuss this issue below. As we have 
stressed above, the convergence of the series means that the
tidal response is mainly determined by a few 'global' modes with
a large scale distribution of $W_{(k)}$ over the planet.

\subsubsection {Degenerate modes with $\sigma_{(k)}=0$ and the case
with $m=0$}
The results of our numerical calculations reported below allow us to
suggest that the numerical method described above is stable and robust
for the modes with sufficiently large  values of 
$|\sigma_{(k)}|$. However, there is a problem with this approach related 
to the case of small values of $|\sigma_{(k)}|\approx 0$. This
problem is associated   with the existence of a non-trivial null space of the 
operator $\bmth{ C}$. It is easy to see that all modes satisfying
\begin{equation} 
\bmth{ C}W_{(k)}=0
\label{eq p87}
\end{equation} 
are degenerate solutions of $(\ref {eq p45})$ (and $(\ref {eq p47})$) with
$\sigma_{(k)}=0$. In fact any $W_{(k)}$ that is independent
of $z$ will do. The corresponding norm $n_{(k)}=0$.
Physically, for $m=0,$ these modes correspond to introducing an infinitesimal
amount of differential rotation and thus slightly shifting the
static equilibrium.
As follows from equation $(\ref {eq p64})$ the
overlap integrals are formally infinite for these modes. However,
in the exact theory this  does not lead to any difficulty since the
contribution of a particular mode to the energy exchange is 
proportional to $\sigma_{(k)}^{4}$ (see e.g. equation   
$(\ref {eq p62})$.  This means that 
the degenerate set of modes is in fact not excited.

However, due to numerical inaccuracy of our method the modes 
formally belonging to the null
space of $\bmth{ C}$ acquire small non zero $\sigma_{(k)}$. The 
corresponding overlap integrals are very large and they can lead, in
principal, to a very large unphysical energy transfer. 
This problem is not significant for the case of $m=2$. In this case,
the quantities $I_{2,-2}(y)$ entering in $(\ref {eq p62})$ are
very small for an interesting range of $y$. They suppress effectively
the contribution of the spurious modes to the energy and angular 
momentum exchange. On the other hand, in the case of $m=0$ 
the corresponding quantities  $I_{2,0}(y)$ do not suppress this
contribution, and the spurious modes do lead to a very large
unphysical energy transfer. Therefore, in principle, the numerical
method must be able to overcome this difficulty, by e.g. incorporating
the solutions to $(\ref {eq p87})$ in the set of the trial functions.

This can be done by considering polynomials consisting
of  products of different power of 
$\varpi$ and $z$ as a new set of trial functions (PP) instead of
solutions to $(\ref {eq p85})$. However, this set is difficult to
use when considering a large values of $j_{max}$. Therefore, in 
our numerical work we simply identified the spurious modes
corresponding to $m=0$ by comparing
the  spectra calculated using these two different sets of 
trial functions and removed them from the summation  
in the expression for the transfer of energy 
(there is no angular momentum exchange in this case).
The resulting contribution to the energy exchange
appears to be very small 
compared with what is obtained for the case of $m=2$, 
and we will not discuss further the case $m=0$ below.    
\subsection{  The spectrum of a planet with $M_*=1M_J$  and $R=1R_J$}

\begin{figure}
\begin{center}
\vspace{8cm}\includegraphics{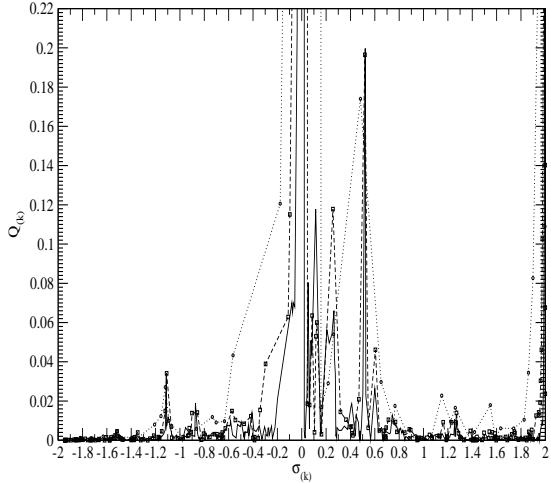}
\end{center}
\vspace{-1cm}
\caption{The overlap integrals $Q_{(k)}$ as functions of the eigen
frequency $\bar \sigma_{(k)}$. The dotted and  dashed  
curves correspond  to $j_{max}=25$ and $j_{max}=100$  respectively.
 Circles and  squares  show 
the positions of particular $Q_{(k)}$. The solid curve  corresponds
to the case  with
$j_{max}=225$.}
\label{Fig.6}
\end{figure}

\begin{figure}
\begin{center}
\vspace{8cm}\includegraphics{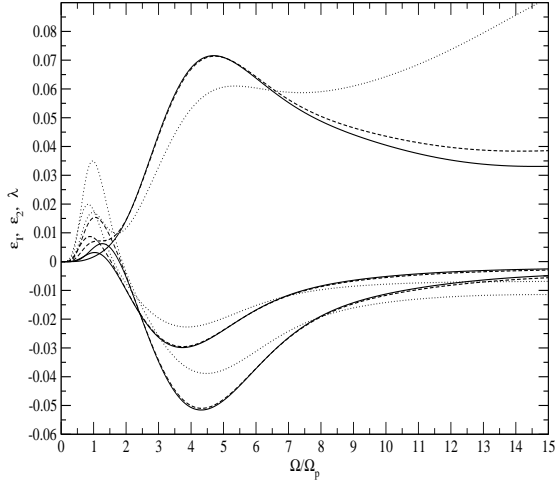}
\end{center}
\vspace{-1cm}
\caption{The quantities $\epsilon_{I}$, $\epsilon_{2}$ and $\lambda$
as functions of the dimensionless rotation rate $\bar \Omega$. The
dotted, dashed and solid curves correspond to $j_{max}=25$, $100$ and
$225$, respectively. The curves which always pass through positive values 
 represent $\epsilon_{2}$. The curve  of a given type
 that takes on the least value corresponds to
 $\epsilon_{I},$ the remaining curve corresponding to $\lambda$.}
\label{Fig.7}
\end{figure}

\begin{figure}
\begin{center}
\vspace{8cm}\includegraphics{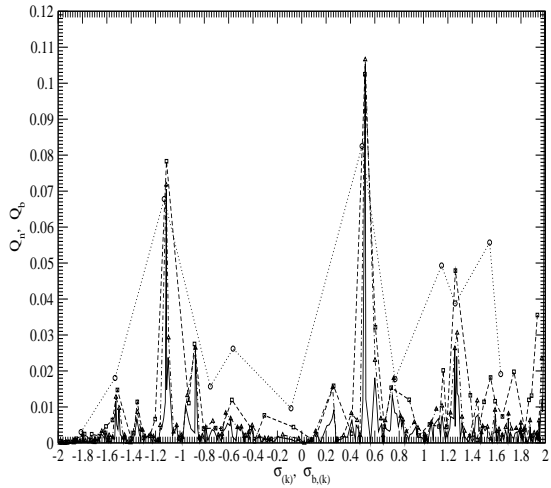}
\end{center}
\vspace{-1cm}
\caption{The binned overlap integrals $Q_{b,(k)}$ are   plotted for
$j_{max}=25$ using the dotted curve   with circles;
 for $j_{max}=100$  using the
dashed curve  with squares;  and for $j_{max}=225$   using the dot-dashed curve
with triangles.
The solid curve shows the original overlap integrals
$Q_{n,(k)}$ calculated for the case of $j_{max}=225$. Note that the
dot-dashed and the solid curves almost coincide.}
\label{Fig.8}
\end{figure}

\begin{figure}
\begin{center}
\vspace{8cm}\includegraphics{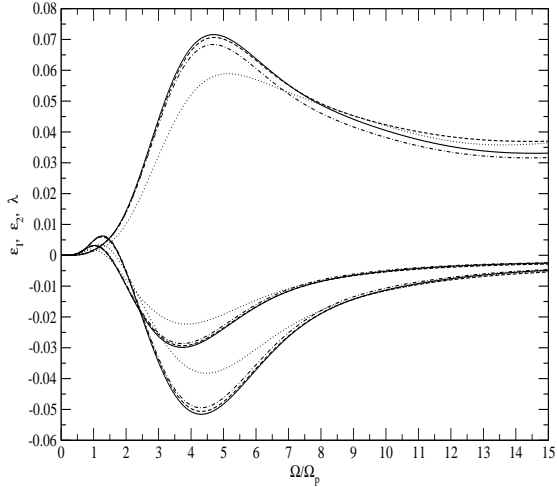}
\end{center}
\vspace{-1cm}
\caption{Same as Fig. \ref{Fig.7} but now the dotted,
 dashed and dot-dashed curves
are calculated   using the binned overlap integrals for
$j_{max}=25$, $j_{max}=100$ and $j_{max}=225$, respectively. The solid
curve is calculated   using the original overlap integrals for
the case of $j_{max}=225$.}
\label{Fig.9}
\end{figure}

\begin{figure}
\begin{center}
\vspace{8cm}\includegraphics{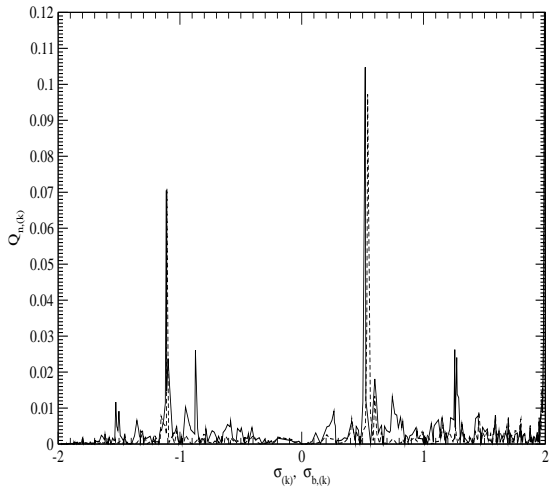}
\end{center}
\vspace{-1cm}
\caption{The overlap integrals $Q_{n,k}$ are shown for two models of
the planet: the model without the phase transition - dashed curve,
and the original model with the phase transition -  solid curve.}
\label{Fig.10}
\end{figure}

\begin{figure}
\begin{center}
\vspace{8cm}\includegraphics{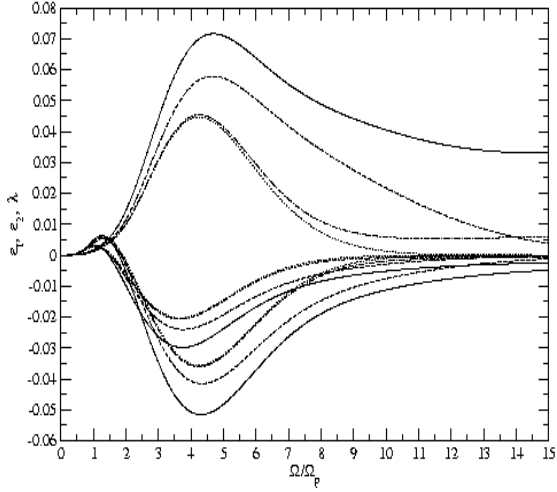}
\end{center}
\caption{Same as Figures \ref{Fig.7} and \ref{Fig.9}, 
but now the dotted and dashed curves
are calculated with   using the binned overlap integrals corresponding
to the global modes. We take into account only the two 'main' global
modes to calculate the dotted curve and the eight global modes
discussed in the text to calculate the dashed curve. The dot-dashed
curve is calculated   using the overlap integrals corresponding
to the model without the phase transition. Note the dotted and
dot-dashed curves for $\epsilon_{I}$ and $\lambda $ practically coincide.}
\label{Fig.11}
\end{figure}

\begin{figure}
\begin{center}
\vspace{3.5cm}
\vspace{7cm}\includegraphics{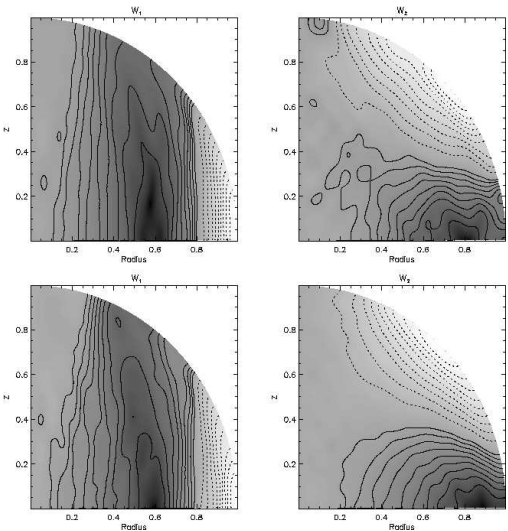}
\end{center}
\caption{The distribution of $W_{(k)}$ over the planet. $W_{(k)}=0$ at
the origin. Contour lines of different style correspond to values of
$W_{k}$ of opposite signs (which are always found to occur). Ten 
contour levels of $W_{n}$ (n=1,10)   for 
 each  sign are defined as follows: 
$W_{n}={1\over n}W_{*}$, where $W_{*}$ is
 one of the  maximum or minimum value of $W_{(k)}.$
 The upper left (right) plot corresponds to
the main global mode $W_{1}$ with $\sigma_{1}=0.521$ ($W_{2}$ with
$\sigma_{2}=-1.11$). The lower left and right plot show the same
distributions but calculated for the planet model without 
the phase transition.}
\label{Fig.12}
\end{figure}

\begin{figure}
\begin{center}
\vspace{3.5cm}
\vspace{7cm}\includegraphics{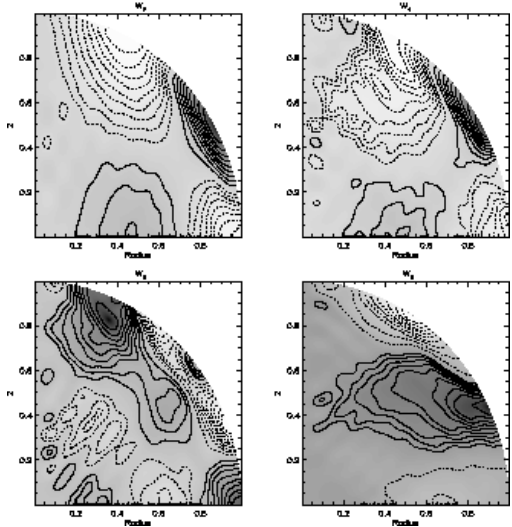}
\end{center}
\caption{Same as Fig. \ref{Fig.12}  but  for  $W_{3}$
with $\sigma_{3}=0.873$ ( upper left plot); $W_{4}$
with $\sigma_{4}=1.257$ ( upper right plot); $W_{5}$
with $\sigma_{5}=1.273$ ( lower left plot); $W_{6}$
with $\sigma_{6}=-1.525$ ( lower right plot).}
\label{Fig.13}
\end{figure}

\begin{figure}
\begin{center}
\vspace{3.5cm}
\vspace{7cm}\includegraphics{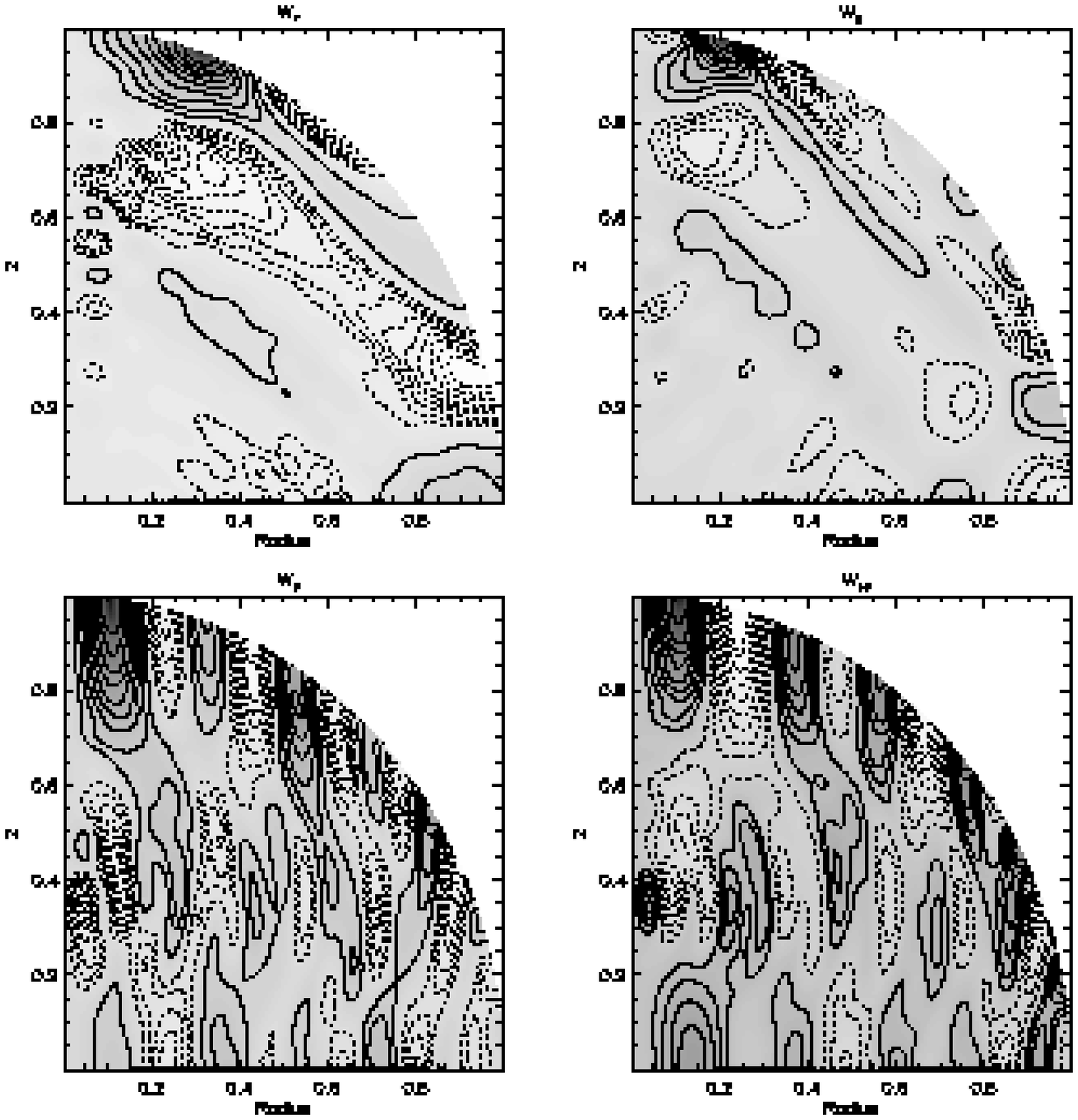}
\end{center}
\caption{ As in Fig. \ref{Fig.12}
but here we show the distributions of $W_{7}$ with $\sigma_{7}=-1.5$
in the upper left plot and $W_{8}$ with $\sigma_{8}=-1.353$ in the
upper right plot. Two non-global modes showing small
scale structure  with $\sigma_{9}=0.49$ 
(lower left plot) and  $\sigma_{10}=-0.5$ 
(lower right plot), are also shown.}
\label{Fig.14}
\end{figure}

In this Section we discuss in detail the case of a planet with
$M_{*}=1M_{J}$ and $R_{*}=1R_{J}$. This case has the most
complicated spectrum of all models we consider. In our opinion,
this is mainly because of the presence of the phase transition 
at a sufficiently small radius $r\approx 0.82$.
The analysis of the eigenfrequencies and the eigenmodes developed
for this case can be easily applied to planets with larger radii
and masses which have a simpler structure.

In Figure \ref{Fig.6}
 we show the overlap integrals $Q_{(k)}$ (see equation $(\ref {eq
p64})$ for the definition) calculated for this case as a function of the
eigenfrequency $\bar \sigma_{(k)}$. It can be shown analytically (PP)
that  $|\bar \sigma_{(k)}| < 2$. The curves of different types
correspond to different numbers of the trial functions, $j_{max}=25$, 
$100$ and $225$. We show
positions of particular  $Q_{(k)}$ representing results of 
calculations with $j_{max}=25$ and $j_{max}=100$ as circles and
squares, respectively. As we have mentioned above the sharp rise near
$\bar \sigma_{(k)} \sim 0$ is due to numerical inaccuracy.
There is a similar unphysical rise near $\bar \sigma_{(k)}
\sim 2$. It is clearly seen from  Figure \ref{Fig.6} that the width of these
numerical features depends on the number of  trial functions 
$j_{max}$ decreasing with increase of $j_{max}$. Similar to the 
case of a polytropic star with index $1.5$
 (PI), there are two 'global' modes with
large values of the overlap integrals, with 
$\bar \sigma_{(1)}\approx 0.5$ and $\bar \sigma_{(2)}\approx -1.1$. The
  frequencies and amplitudes of these modes depends only weakly  on
the number of  trial functions, and these modes are especially 
important for the problem on hand.  However, contrary to 
the case of a polytropic star with index $1.5$,
 there are some other markedly large values 
of $Q_{(k)}$ associated with modes at other eigenfrequencies (eg. near 
$\sigma_{3}\approx -0.9$ and $\sigma_{4}\approx 1.25$). 
These can also be associated with
global modes, which are, however, less stable with respect 
to  changes of $j_{max}$. They cannot be
reliably identified with a small number of the trial functions
($j_{max}=25$ in our case). We discuss properties of these modes
later in this section.

In Figure \ref{Fig.7}  we show the quantities
$\epsilon_{2},$  
$\lambda$   
and
$\epsilon_{I}$ defined by equations 
(\ref{eq p891}-\ref{eq p89})  
which determine the energy and angular momentum transfer through
equations  $(\ref {eq p65})$
and $(\ref {eq p66})$.  
As for figure \ref{Fig.6} the curves of different types correspond to 
different values of $j_{max}$. It is seen from this Figure that
when $\bar \Omega $ is sufficiently large the curves representing
large values of $j_{max}=100$ and $225$ are in a very good agreement 
with each other. The curves representing $\epsilon_{I}$ and $\lambda$ for 
$j_{max}=25$ are also in qualitative agreement with those
corresponding to the larger value of $j_{max}$. However, the
curve representing $\epsilon_{2}$ deviates significantly when 
$\bar \Omega > 7$. When $\bar \Omega $ is small, the curves
representing $j_{max}=25$ and $100$ have larger values compared
to the case with $j_{max}=225$. This is due the contribution of 
spurious modes with  eigenfrequencies very close to $\sigma =2$.
In principle, we can improve the agreement between different curves 
significantly simply by discarding these modes. However, as we will 
see below it is possible to 
improve the stability of our numerical scheme with respect
to the choice of different numbers of  trial functions and 
suppress the influence of these spurious modes automatically.     

\subsubsection{Binned eigenfrequencies and overlap integrals}
In order to improve the stability of our numerical scheme with respect to
numerical noise we would like to
separate the eigenspectrum into  bins  with a finite frequency width and 
 adopt  the sum of the overlap integrals within a particular bin instead of
$Q_{(k)}$. To begin, we redefine the
overlap integral according to the rule
\begin{equation}  
Q_{n, (k)}=\bar \sigma_{(k)}^{2}\sqrt {(4-\bar \sigma_{(k)}^2)}Q_{(k)}. 
\label{eq p91}
\end{equation}
It follows from equation  $(\ref {eq p891})$, that  the quantity
$\epsilon_{2}$ has especially simple form when expressed in terms
$Q_{n, (k)}$, namely
\begin{equation} 
\epsilon_{2}=C_2\bar \Omega^4\sum_{(k)}  
Q_{n,(k)}^{2}I^{2}_{2,-m}(y_k).
\label{eq p92}
\end{equation}   
To define the frequency bins
let us separate the eigenspectrum   into regions  between  successive  
minima of $Q_{n,(k)}^{2}$ and   treat all $Q_{n,(k)}$
and $\sigma_{(k)}$ within  such a region 
as belonging to the same bin. Note that modes defining minima
are not included in the bins. Accordingly, we have
\begin{equation}
Q_{b,(l)}=\sqrt {\sum Q_{n,(k)}^{2}}, \quad 
\sigma_{b,(l)}={(\sum Q_{n,(k)}^{2} \sigma_{(k)})\over Q^{2}_{b,(l)}},
\label{eq p93}
\end{equation}  
as new 'binned' overlap integrals and 'binned' eigenfrequencies.
 Here, the summation is performed over all eigenfrequencies
belonging to the same bin, and $l$ is a number of a bin
\footnote{Note that this procedure is not unique, and 
better results may be achieved with a more sophisticated scheme of
binning. However, our procedure is, perhaps, the simplest one, and it
is definitely sufficient for our purposes.}. 
 
The result of comparison between the 'binned' quantities  $Q_{b,k}$
corresponding to different $j_{max}$
and the quantities $Q_{n,k}$ calculated with the largest
$j_{max}=225$ is shown in Figure \ref{Fig.8}. One can see from this Figure that
the binned quantities calculated for $j_{max}=100$ and $j_{max}=225$
are in a good agreement. The results corresponding to $j_{max}=25$
agree with what is shown for larger values of $j_{max}$ for the two
'main' global modes. They are also in qualitative agreement for 
 $\sigma_{(k)} < 0$, but deviate in a part
of the spectrum with positive values of $\sigma_{(k)}$. It turns out
that this deviation is not important for our purposes. Indeed, 
the positive part of the spectrum plays a role in the tidal response
problem when $\bar \Omega$ is small. In this case width of 
the function $I_{2,-2}(y)$ is rather large, 
and the tidal response is mainly determined by the 'main' global mode with 
$\bar \sigma_{1}\approx 0.5$ and the largest overlap integral. 
When $\bar \Omega$ is large, the tidal response is mainly determined 
by the modes with negative $\sigma_{(k)}$ which can then comove
with the tidal perturbation near periastron. 
The width of $I_{2,-2}(y)$ decreases with increase of $\bar \Omega$,
and, when the rotation rate is large, the tidal response is determined 
by several modes  with sufficiently large overlap integral and
sufficiently large absolute values of $\sigma_{(k)}$. This effect
explains in part why the energy transfer in the rotating frame 
determined by the quantity $\epsilon_{2}$ tends to some non-zero 
value with increase of the rotation rate (see e.g. Figure 7 and Figure
9).  

In Figure \ref{Fig.9} 
 we show the quantities $\epsilon_{I}$, $\epsilon_{2}$ and 
$\lambda $ calculated   using the binned overlap integrals 
for the different values of $j_{max}$ and 
the corresponding quantities calculated   using the non-binned
overlap integrals with $j_{max}=225$. Comparing Figure \ref{Fig.7}
with Figure \ref{Fig.8}
we see that the use of the binned quantities does allows us to improve
significantly the stability of our numerical scheme with respect to
different numbers of basis functions. Now, all curves are in a 
good agreement with each other for the whole considered 
range of $\bar \Omega$. 

\subsubsection{Eigenspectrum of the model without the phase
transition}
In Figure \ref{Fig.10} we show the result of calculation of the overlap
integrals for the model with 'smoothed out' phase transition (dashed
curve) together with the overlap integrals for the model with 
the phase transition. As it is seen from this Figure the two 
'main' global modes have practically the same overlap integrals for
the both models. However, other prominent 
spikes present in the model with a phase transition are absent for the 
model without the phase transition. Therefore, there
only two global modes in that case. One can assume that
the presence of the other global 'non standard' modes in the realistic
planet models is related to the presence of 
the phase transition. In fact, as we will see below these modes are
also present in the planet models with larger radii
where the phase transition is absent.
It is reasonable to suppose that the presence of some of
these  modes is related to a sharp decrease of the density
after the radius of the
phase transition (or the melting point for the models with larger
radii, see Figures \ref{Fig.3}  and \ref{Fig.4})
in the realistic planet models.    
\subsubsection{Global modes} 
As we have pointed out above there is a
set of modes characterised by relatively large values of the overlap
integrals. These modes are referred to as global
modes. Physically, these modes have  eigenfunctions   for which
 $W_{(k)}$ varies on a large scale. In this
situation, the overlap integrals are not averaged to small values after
integration over the volume of the planet. 

As  can be seen from Figures \ref{Fig.6}, \ref{Fig.8}
 and \ref{Fig.10} there are more than ten modes with 
markedly large overlap integrals. However, they are not equally
important for the problem on hand. 
The two most important modes are 'the main' global modes with the largest
values of the overlap integrals.
These are present in all our models as well as in polytropic stars. 
Other global modes 
with negative values of $\sigma_{(k)}$ can also play an important 
role in the tidal response. Additionally, 
it is interesting to consider  the positive and 
negative frequency global modes with the largest  values 
of $Q_{n,(k)}$ after the main modes. 
These modes contain information about the
difference of the realistic planet models from over-simplified polytropes,
and may be of interest for another problems as well. There are eight modes
satisfying these criteria. Their eigen frequencies and overlap
integrals are shown in Table \ref{table1}. 
\begin{table*}
 \centering
 \begin{minipage}{140mm}
  \caption{ \label{table1}
  The eigen frequencies and overlap integrals of the global modes.}
  \begin{tabular}{@{}llrrrrlrlr@{}}
  \hline
   $W_{(k)}$ & $\sigma_{(k)}$ & $\sigma_{b,(k)}$ & $Q_{n,(k)}$ & $Q_{b,(k)}$\\
      
 \hline
   $W_{1}$ & 0.521 & 0.521 & 0.105 & 0.1065 \\  
   $W_{2}$ & -1.11 & -1.11 & 0.07 & 0.072 \\
   $W_{3}$ & -0.873 & -0.872 & 0.026 &0.026 \\
   $W_{4}$ & 1.257 & 1.257 & 0.026 & 0.026 \\
   $W_{5}$ & 1.273 & 1.276 &  0.024 & 0.03 \\
   $W_{6}$ & -1.526 & -1.525 & 0.0117 & 0.0127 \\
   $W_{7}$ & -1.5 & -1.5 & 0.0091 & 0.0096 \\
   $W_{8}$ & -1.353 & -1.349 & 0.0066 & 0.0085 \\

\hline
\end{tabular}
\end{minipage}
\end{table*}
$W_{1}$ and $W_{2}$ are the two 'main' global modes. 
$W_{4}$ and $W_{5}$ as well as $W_{6}$ and $W_{7}$ have very close eigen
frequencies. Most probably, these modes are doublets split as a 
result of a perturbation.

In Figure 11 we show the quantities $\epsilon_{I}$, $\epsilon_{2}$ and 
$\lambda $ calculated with only two global modes (the dotted curves), 
with the eight global modes (the dashed curves), and with the whole
spectrum of modes (the solid curves). The dot-dashed curves represent 
the corresponding quantities calculated for the model without the 
phase transition. One can see that the curves for the model without 
the phase transition are very close to the 
curves calculated  using only  the two 'main' global modes - the
result expected from our previous discussion.  In general, the curves
calculated using the global modes give a good approximation to
the curves calculated with help of the whole spectrum excepting  
the curves representing the transfer of energy in the inertial frame,
$\epsilon_{2}$ at a high rotation rate, $\bar \Omega > 10$. 
The solid curve determined by the whole spectrum 
gives much larger energy transfer in this case. 
This is due to the influence of the modes with eigen frequencies very
close to $-2$. It is not clear whether this effect is physical or it 
is determined by numerical inaccuracies of our method. A calculation
with significantly larger number of the trial functions may resolve 
this question. However, this effect is unimportant for our purposes
and will not be discussed further.

In Figures \ref{Fig.12}-\ref{Fig.14}
 we show the distribution of $W_{k}$ over the planet 
for the eight global modes. Additionally, in Figure \ref{Fig.14} we show
the distribution of $W_{k}$ for two 'non-global' modes: $W_{9}$ with
$\sigma_{9}=0.49$ and $W_{10}$ with $\sigma_{10}=-0.5$. These modes 
have very small overlap integrals. In the   upper panels 
of Figure \ref{Fig.12}
 the two 'main' modes
$W_{1}$ and $W_{2}$ are shown, and in the lower panels of the same Figure
we show the same modes calculated for the model without the  phase
transition. One can see that these modes have essentially the same 
structure for the both models. 
Moreover, these distributions are very similar 
to the distributions of the corresponding modes for the case of  a $n=1.5$
polytrope (IP). Therefore, it is natural to suggest that the
dependence of the structure of the 'main' modes on the structure of 
a planet is rather insignificant (see also Figures \ref{Fig.19} 
and \ref{Fig.23} below). 
Other global modes
have structures with amplitudes  that are  more   concentrated 
towards the boundary of the planet 
when compared to the
'main' modes.  It is instructive to compare the
global and non-global modes. One can see from Figure \ref{Fig.14} that the 
non-global modes are characterised by a large number of patterns with 
 alternating signs. These patterns give compensating contributions to the
overlap integrals thus significantly decreasing their values.

\subsection{Eigenspectra of the planets with larger masses and radii} 
 
\begin{figure}
\begin{center}
\vspace{8cm}\includegraphics{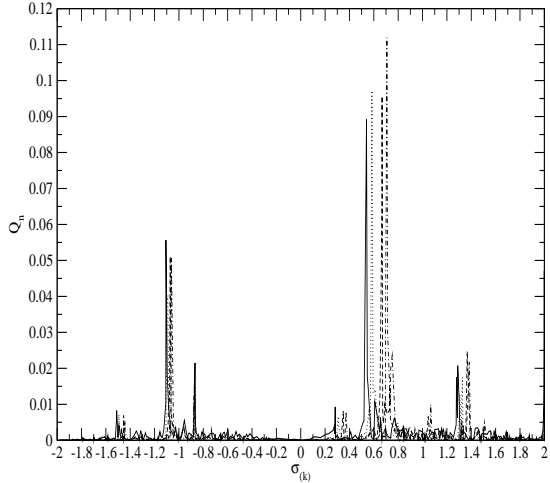}
\end{center}
\vspace{-1cm}
\caption{The overlap integrals $Q_{n,(k)}$ are shown for
the planet models with $M_{*}=1M_{J}$, and four different radii:
$R_{*}=1.2R_{J}$ (solid curve), $R_{*}=1.4R_{J}$ ( dotted curve),
$R_{*}=1.8R_{J}$ (dashed curve), and $R_{*}=2R_{J}$ 
( dot-dashed curve).}
\label{Fig.15}
\end{figure}

\begin{figure}
\begin{center}
\vspace{8cm}\includegraphics{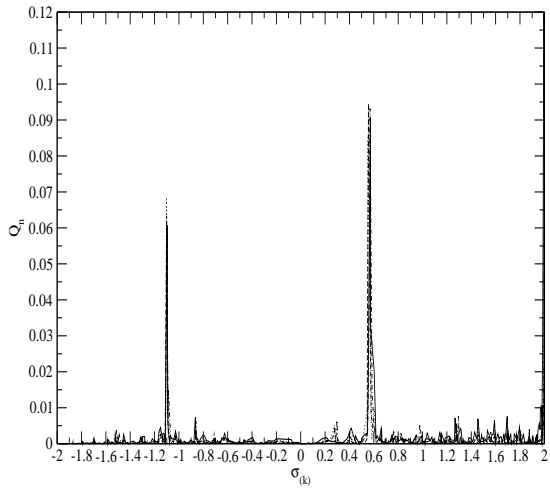}
\end{center}
\vspace{-1cm}
\caption{The same as Fig. \ref{Fig.15}
 but for $M_{*}=5M_{J}$. The solid, dotted,
dashed and dot-dashed curves are for $R_{*}=1.03R_{J}$, $1.4R_{J}$,
$1.6R_{J}$ and $2R_{J}$, respectively.}
\label{Fig.16}
\end{figure}

\begin{figure}
\begin{center}
\vspace{8cm}\includegraphics{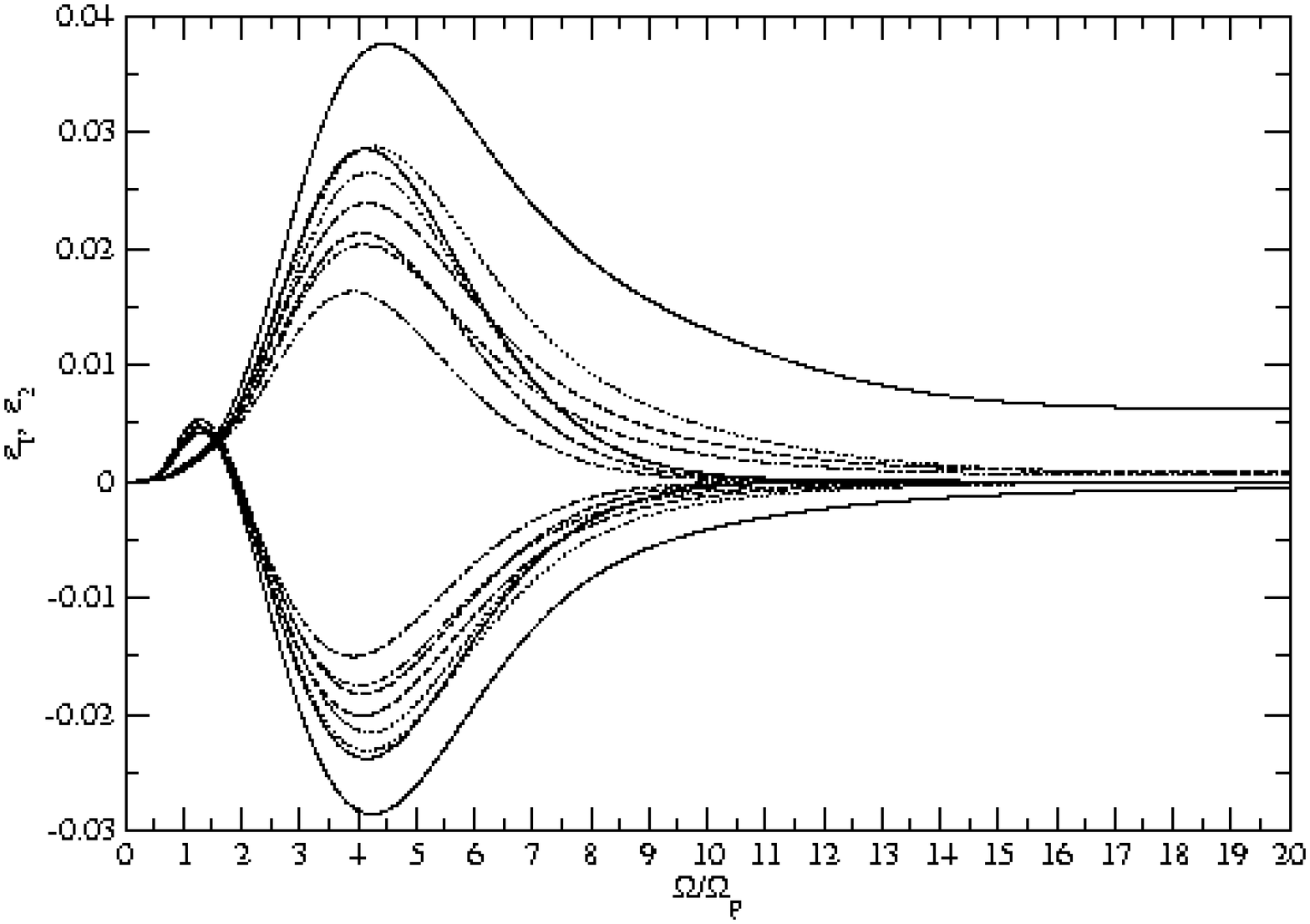}
\end{center}
\caption{The dependence of $\epsilon_{I}$ and  $\epsilon_{2}$ for the
planet models with $M_{J}=1M_{*}$ and 
$R_{*}=1.2R_{J}$ ( solid curves), $R_{*}=1.4R_{J}$ ( dotted curves),
$R_{*}=1.8R_{J}$ ( dashed curves), and $R_{*}=2R_{J}$ 
( dot-dashed curves). The curves which always  pass through positive values
correspond to $\epsilon_{2}$. The curves of the same type with smaller
amplitude  correspond to $\epsilon_{I}$ and 
$\epsilon_{2}$ calculated  using only  the two 'main' global modes.}
\label{Fig.17}
\end{figure}

\begin{figure}
\begin{center}
\vspace{8cm}\includegraphics{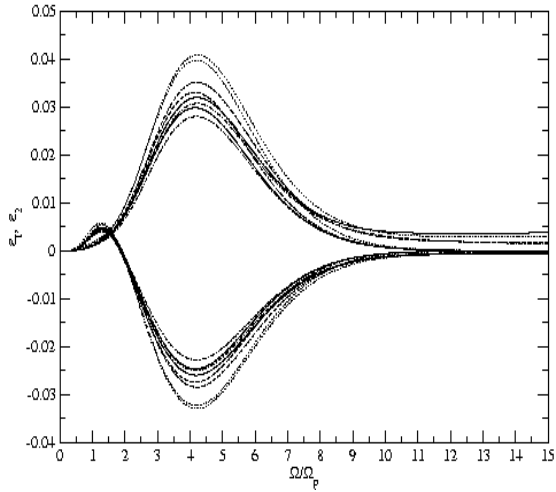}
\end{center}
\caption{The same as Fig. \ref{Fig.17}
 but for $M_{*}=5M_{J}$. The solid, dotted,
dashed and dot-dashed curves are for $R_{*}=1.03R_{J}$, $1.4R_{J}$,
$1.6R_{J}$ and $2R_{J}$, respectively.}
\label{Fig.18}
\end{figure}

In Figure \ref{Fig.15} we show the dependency of the overlap integral
$Q_{n,(k)}$ on the eigenfrequency $\sigma_{(k)}$ for the planet 
models with planet mass equal to the mass of Jupiter   with four different
radii, $R_{*}=1.2R_{J}$, $1.4R_{J}$, $1.8R_{J}$ and $2R_{J}$. 
The general structure of the spectrum is close to that 
of the planet with $M_* = 1M_J,$ and $R_* = 1R_J,$ see ( Figure \ref{Fig.8}). 
In all cases there are the two 'main' global modes. Also there are
some other 'non-standard' global 
modes with  eigenfrequencies close to the corresponding
eigenfrequencies of the  model with $R_* = 1R_J.$ The 'main' modes
slightly change their position with growth of planet radius, and their
overlap integrals also slightly depend on the radius, being however
rather close to each other. It is interesting to point out that the 
overlap integrals corresponding to the 
'non-standard' global modes are 
smaller than those corresponding to the model with  $M_* = 1M_J,$ and
and $R_* = 1R_J.$ Therefore,
the influence of the 'main' global modes on the energy transfer is
more pronounced than in that  case (see Figure \ref{Fig.17}). 

In Figure \ref{Fig.16}  we show the results of
calculations for  models which have  $M_{*}=5M_{J}$ with 
$R_{*}=1.03R_{J}$, $1.4R_{J}$, $1.6R_{J}$ and $2R_{J}$. As seen from 
this Figure the spectrum is quite stable for this case, 
 such that the curves corresponding to different planet radii 
can be hardly distinguished from each other. Also, the spectrum is
totally dominated by  the two 'main' global modes with $\sigma_{1}\approx
0.55$ and $\sigma_{2}\approx -1.1$. In fact, for these
models it is possible to identify reliably only three non-standard
global modes with $\sigma_{(k)}\approx -0.85$, $0.3$ and $1.3$. These
modes, however, have rather small overlap integrals.

In Figures \ref{Fig.17} and \ref{Fig.18}
 we show the dependence of the quantities
$\epsilon_{I}$ and $\epsilon_{2}$ which determine the energy transfer
in the inertial and rotating frames, respectively, 
on $\bar \Omega$ for the set of
planet model considered in this Section. Also, we compare 
these quantities with the same quantities calculated using 
only the two 'main' global modes  and the 'binned' overlap integrals and
 eigenfrequencies discussed above. The corresponding 
eigenfrequencies and overlap integrals are shown in Tables \ref{table2} 
and \ref{table3}. 
\begin{table*}
 \centering
 \begin{minipage}{140mm}
  \caption{\label{table2}The binned eigen frequencies and overlap integrals of the
  main modes for $M_{*}=1M_{J}$.}
  \begin{tabular}{@{}llrrrrlrlr@{}}
  \hline
   $R_{*}/R_{pl}$ & 1.2 & 1.4 & 1.8 & 2\\
      
 \hline
   $\sigma_{b,1}$ & 0.54 & 0.58 & 0.65 & 0.71 \\  
   $Q_{b,1}$ & 0.093 & 0.097 & 0.117 & 0.12 \\
   $\sigma_{b,2}$ & -1.1 & -1.09 &-1.076  &-1.037 \\
   $Q_{b,2}$ & 0.057 & 0.06 & 0.054 & 0.051 \\

\hline
\end{tabular}
\end{minipage}
\end{table*}

\begin{table*}
 \centering
 \begin{minipage}{140mm}
  \caption{\label{table3}The binned eigen frequencies and overlap integrals of the
  main modes for $M_{*}=5M_{J}$.}
  \begin{tabular}{@{}llrrrrlrlr@{}}
  \hline
   $R_{*}/R_{pl}$ & 1.03 & 1.4 & 1.6 & 2\\
      
 \hline
   $\sigma_{b,1}$ & 0.55 & 0.53 & 0.57 & 0.59 \\  
   $Q_{b,1}$ & 0.1 & 0.105 & 0.098 & 0.096 \\
   $\sigma_{b,2}$ & -1.1 & -1.1 &-1.1  &-1.1 \\
   $Q_{b,2}$ & 0.061 & 0.069 & 0.064 & 0.058 \\

\hline
\end{tabular}
\end{minipage}
\end{table*}

In Figure \ref{Fig.17} we show
the results of calculations for the planet models with $M_{*}=1M_{J}$ 
and in Figure \ref{Fig.18} the case of $M_{*}=5M_{J}$ is shown. As seen from
these Figures,  the quantities determined by using only the 
two 'main' modes give a good
approximation for all curves excepting the cases of the planet models 
that have  $M_{*}=1M_{J}$  with $R_{*}=1.2R_{J}$ and $M_{*}=5M_{J}$ with   
$R_{*}=1.03R_{J}.$ In those cases
 the curves giving $\epsilon_{2}$ deviate 
significantly when $\bar \Omega $ is large. 

\begin{figure}
\begin{center}
\vspace{3.5cm}
\vspace{7cm}\includegraphics{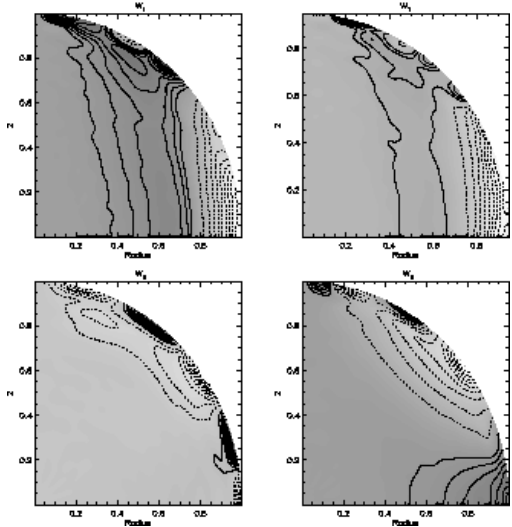}
\end{center}
\caption{The distribution of $W_{(k)}$  for
$M_{*}=1M_{J}$  with two different radii $R_{*}=1.4R_{J}$ and
$R_{*}=2R_{J}$. The two 'main' global modes $W_{1}$ ( upper panels) 
and $W_{2}$ ( lower panels) are 
shown with the left (right) plots corresponding to $R_{*}=1.4R_{J}$ 
($R_{*}=2R_{J}$).
The upper left (right) plot corresponds to
$\sigma_{1}=0.58$ with $Q_{n,1}=0.097$ ($\sigma_{1}=0.71$ 
with $Q_{n,1}=0.11$). The lower left (right) plot corresponds to
$\sigma_{2}=-1.1$ with $Q_{n,2}=0.038$ ($\sigma_{2}=-1.06$ 
with $Q_{n,2}=0.051$).}
\label{Fig.19}
\end{figure}

\begin{figure}
\begin{center}
\vspace{3.5cm}
\vspace{7cm}\includegraphics{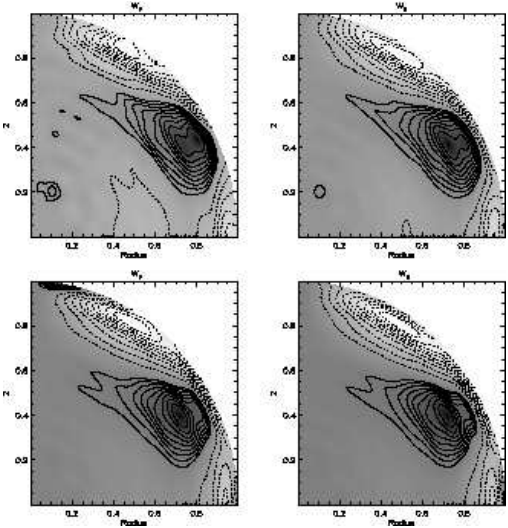}
\end{center}
\caption{The  form of the 'non standard' global mode  
$W_{3}$  with $\sigma_3 \approx -0.9$   for the planet 
models with $M_{*}=1M_{J}$ and four different radii: $R_{*}=1.2R_{J}$,
 upper left plot; $1.4R_{J}$,
 upper right plot$; R_{*}=1.8R_{J}$
 lower left plot; $R_{*}=2R_{J}$,
 lower right plot.}
\label{Fig.20}
\end{figure}

\begin{figure}
\begin{center}
\vspace{3.5cm}
\vspace{7cm}\includegraphics{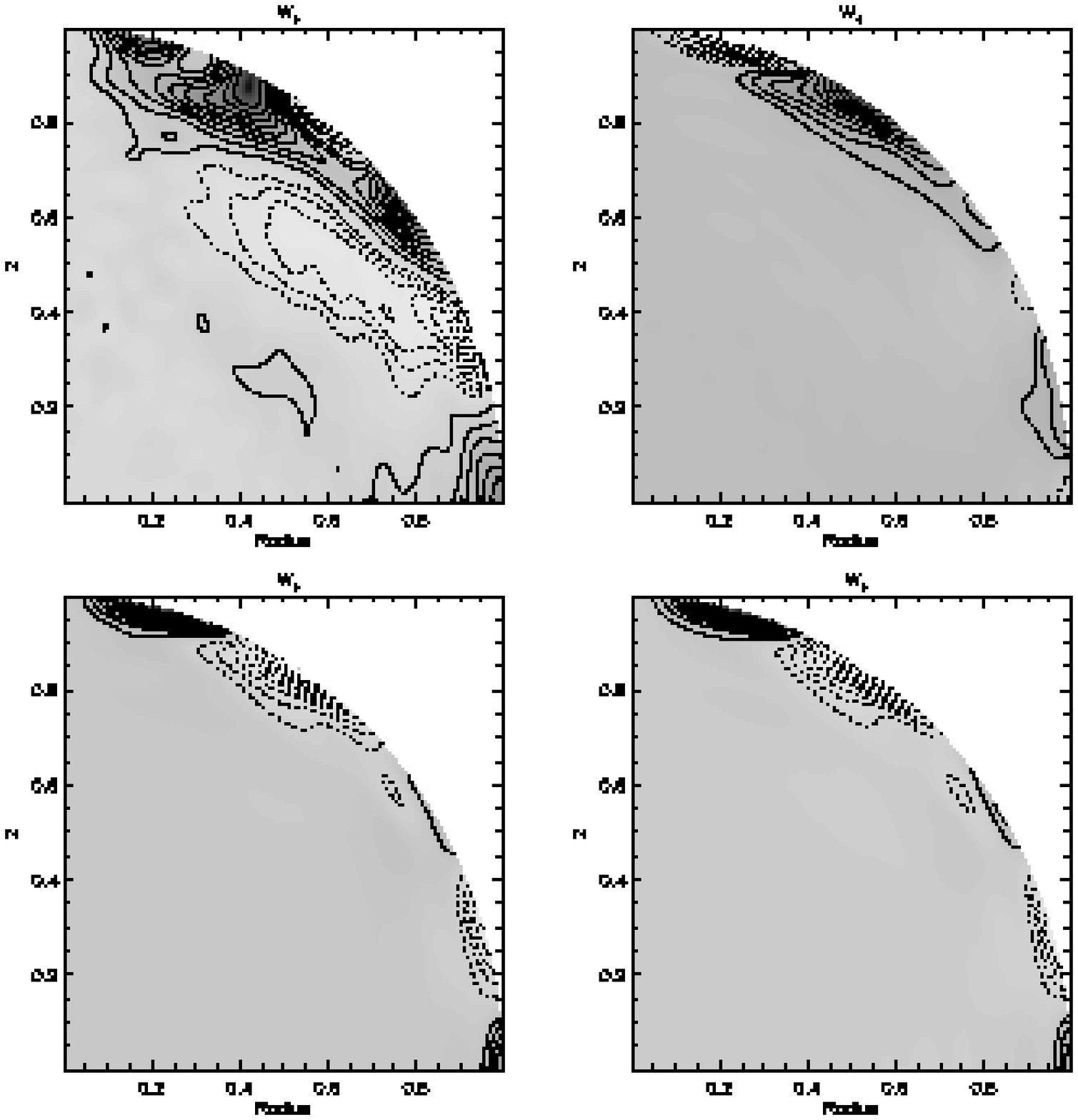}
\end{center}
\caption{Same as Fig. \ref{Fig.20}  but for the global mode
$W_{4}$ with $\sigma_{4}\approx 1.3$.}
\label{Fig.21}
\end{figure}

\begin{figure}
\begin{center}
\vspace{3.5cm}
\vspace{7cm}\includegraphics{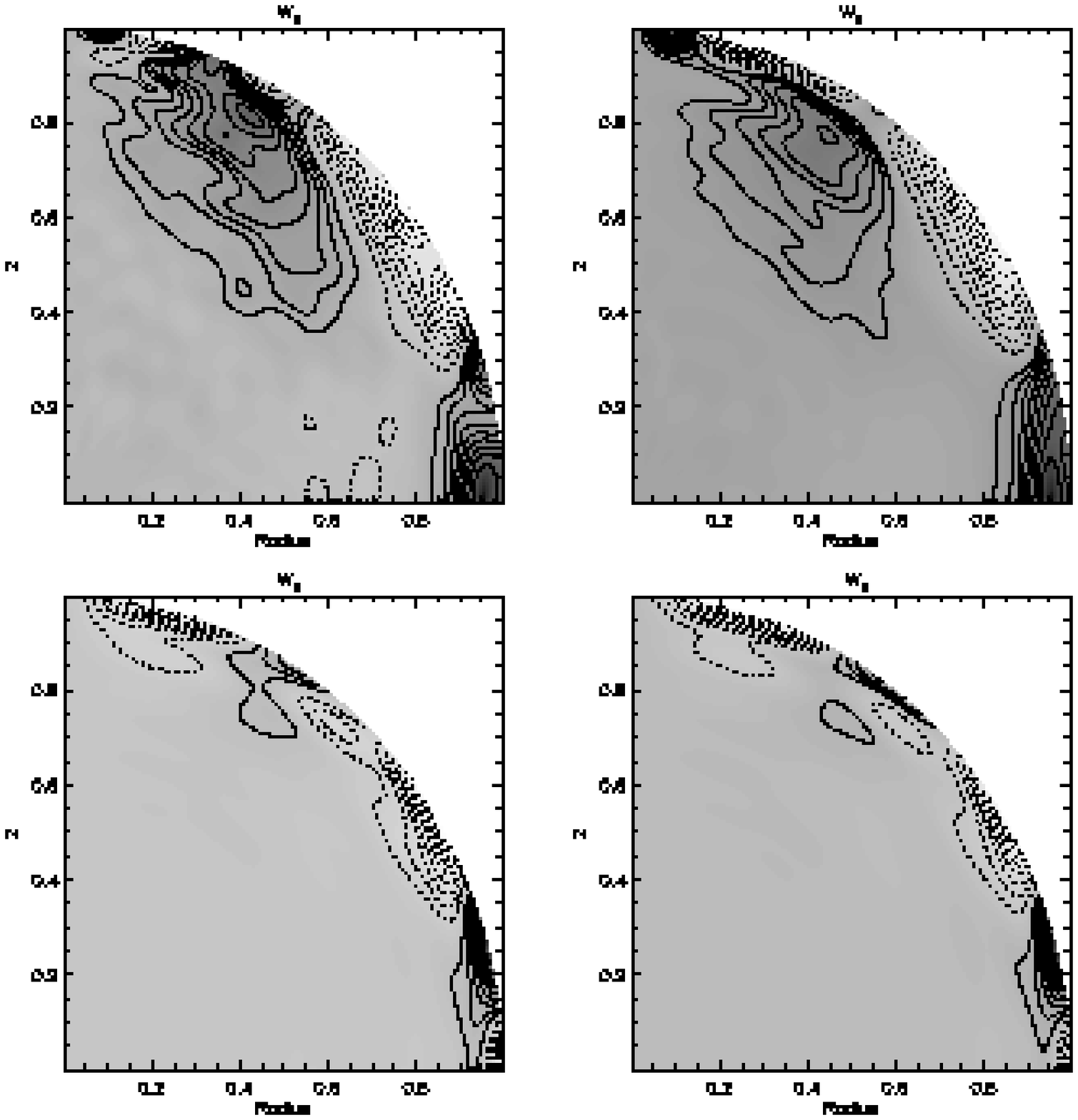}
\end{center}
\caption{Same as Fig. \ref{Fig.20}  
but for the global mode $W_{5}$ 
with $\sigma_{5}\approx -1.5$.}
\label{Fig.22}
\end{figure}

\begin{figure}
\begin{center}
\vspace{3.5cm}
\vspace{7cm}\includegraphics{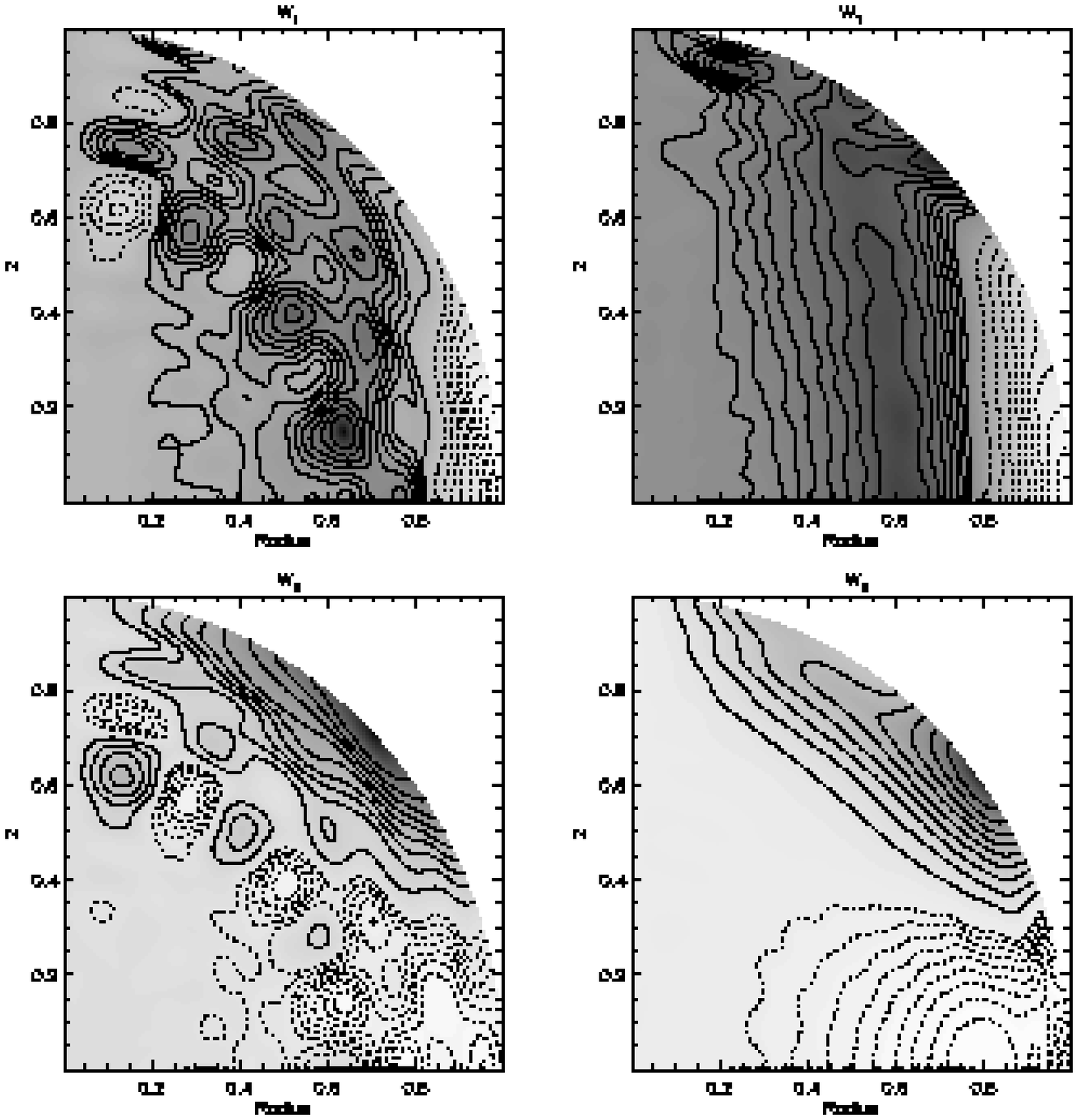}
\end{center}
\caption{The same as Fig. \ref{Fig.19} but for $M_{*}=5M_{J}$.  
The two 'main' global modes $W_{1}$ (upper panels) 
and $W_{2}$ ( lower panels) are 
shown with the left (right) plots   corresponding to $R_{*}=1.03R_{J}$ 
($R_{*}=2R_{J}$).
 For the upper plots 
$\sigma_{1}\approx 0.57$ and $Q_{n,1}\approx 0.09$ 
For the lower plots
$\sigma_{2}\approx -1.1$ and $Q_{n,2}\approx 0.06$, 
for the both models.}
\label{Fig.23}
\end{figure}

Figures \ref{Fig.19}-\ref{Fig.23} illustrate the effects 
of changing planet radius and mass on the form of $W_{(k)}$, 
for the global modes. 

In Figure \ref{Fig.19} we show the distributions  
for the two 'main' global modes $W_{1}$ and $W_{2}$ 
in the case of the planets with
$M_{*}= 1M_{J}$ and  radii larger than the radius of
Jupiter. Comparing this Figure and Figure \ref{Fig.12} we see that the spatial
structure of the 'main' global modes only slightly changes with change
of planet radius. These modes may be considered as 'stable' modes with
respect to  changes of this parameter. In Figure \ref{Fig.20} we show the
influence of changing  radius on the 'non standard' global 
mode $W_{3}$ with $\sigma_{3} \approx -0.9$. The distribution of this
mode is practically the same for  planet radii in the range 
$1.2R_{J}-2R_{J}$, and therefore this mode is particularly 'stable'.
Examples of 'non-stable' modes are shown in Figures \ref{Fig.21}
 and \ref{Fig.22}. We
  plot the mode $W_{4}$ with $\sigma_4\approx 1.3$ in Figure \ref{Fig.21}
and the mode $W_{5}$ with $\sigma_5 \approx -1.5$ in Figure \ref{Fig.22}.
 One can
see from these Figures that the region  with non-negligible
 $W_{(k)}$ is gets smaller with radius. Accordingly,
the overlap integrals  get smaller as well, see Figure \ref{Fig.15}. 

As follows from Figure \ref{Fig.16} the effective spectra of the models with
$M_{*}=5M_{J}$ is pretty stable with respect to a change of 
radius, and the overlap integrals corresponding to the 'non standard'
global modes are small compared with those corresponding to the
'main' global modes. Therefore, in Figure \ref{Fig.23} we 
show only the  two 'main' global modes 
  for $R_{*}=1.03R_{J}$ and $R_{*}=2R_{J}$.   The general
structure of the modes is similar to the case $M_{*}=1M_{J}$ (see
Figures \ref{Fig.12} and \ref{Fig.19}). 
However, there is an interesting feature of the 
distributions  in the case $R_{*}=1.03R_{J}$. In this 
case a small scale pattern is superimposed on 
the large scale distributions of $W_{1}$ and $W_{2}$. Most probably, 
this effect can be explained by mixing between these two global modes
and  neighbouring local modes as a result of a perturbation.

\section{Rotation rate and energy exchange in the state of pseudo-synchronisation}

When the state of pseudo synchronisation is maintained with the value of
the angular velocity being equal to its equilibrium value
$\Omega_{eq}$, the angular momentum transfer is absent 
$\lambda(\Omega_{eq})=0$. In this case the energy transfer in the
inertial frame is equal to the energy transfer in the rotating frame,
and therefore the corresponding curves in Figures \ref{Fig.9}, \ref{Fig.11},
 \ref{Fig.17} and \ref{Fig.18} 
intersect each other. One can see from these Figures that the
intersection point have similar coordinates for all models we
consider, and therefore the dimensionless 
rotation rate $\bar \Omega_{eq}$ and the quantity $\epsilon_{*}$
determining the energy exchange through equation $(\ref {eq p68})$
only slightly depend on the parameters of the planet models.
We summarise results of calculations of these quantities in Tables \ref{table4} 
and \ref{table5}.

\begin{table*}
 \centering
 \begin{minipage}{140mm}
  \caption{\label{table4} $\epsilon_{*}$ and $\bar \Omega_{eq}$ for the planet models
  with $M_{*}=M_{J}$.}
  \begin{tabular}{@{}llrrrrlrlr@{}}
  \hline
   $R_{*}/R_{J}$ & 1 & 1.2 & 1.4 & 1.6 & 1.8 & 2\\
      
 \hline
   $\epsilon_{*}$ & $5.3\cdot 10^{-3}$ & $3.6\cdot 10^{-3}$ &
   $3.45 \cdot 10^{-3}$  & $3.6 \cdot 10^{-3}$ & $3.7\cdot 10^{-3}$&
   $3.6\cdot 10^{-3}$ \\  
   $\bar \Omega_{eq}$ & $1.46$ & $1.52$ & $1.56$ & $1.59$ & 
   $1.6$ & $1.61$ \\
   \hline
\end{tabular}
\end{minipage}
\end{table*} 

\begin{table*}
 \centering
 \begin{minipage}{140mm}
  \caption{\label{table5} $\epsilon_{*}$ and $\bar \Omega_{eq}$ for the planet models
  with $M_{*}=5M_{J}$.}
  \begin{tabular}{@{}llrrrrlrlr@{}}
  \hline
   $R_{*}/R_{J}$ & 1.03 & 1.2 & 1.4 & 1.6 & 1.8 & 2\\
      
 \hline
   $\epsilon_{*}$ & $3.6\cdot 10^{-3}$ & $3.6\cdot 10^{-3}$ &
   $4.2 \cdot 10^{-3}$  & $3.7 \cdot 10^{-3}$ & $3.55\cdot 10^{-3}$&
   $3.4\cdot 10^{-3}$ \\  
   $\bar \Omega_{eq}$ & $1.55$ & $1.55$ & $1.54$ & $1.54$ & 
   $1.54$ & $1.55$ \\
   \hline
\end{tabular}
\end{minipage}
\end{table*} 

As follows from these table apart from the case of a planet with
Jupiter mass and radius which gives a rather large value 
$\epsilon_{*}=5.3\cdot 10^{-3}$, the values of $\epsilon_{*}$ and
$\bar \Omega_{eq}$ are rather close to each other. Therefore,
in our estimates of the time scale of tidal circularisation below
we will use the 'typical' values $\epsilon_{*}=3.6\cdot 10^{-3}$ 
and $\bar \Omega_{eq}=1.55$. Accordingly, the energy transfer will 
be given by 
\begin{equation}
E_{ps}={3.6\cdot 10^{-3}\over \eta^{6}}E_{*},
\label{eq p94}
\end{equation}  
regardless of the planet radius and mass.

In Figure \ref{Fig.16} we show the 
dependence of $E_{ps}$ on $\eta$ as a solid curve. The dotted, dashed and dot-dashed curves correspond to
the energy transfer associated with the fundamental modes, $E_{f}$. They are calculated with help of
equations derived in  IP and presented in Appendix 
\footnote{ Note a misprint made in (IP). Their equations (60), (61) and (64)
must by multiplied by factor $\pi^2$. This 
has no influence on the conclusions of IP. The correct expressions are shown in Appendix.}. 
In a similar manner to the energy transfer associated with the inertial modes, $E_{f}$ depends on the 
angular velocity
of the planet, and we present several different curves calculated for different values of $\Omega $. 
The dashed curves are calculated for
a non-rotating planet with $\bar \Omega =0$, they correspond to a maximal energy transfer $E_{f}^{max}$
as a function of $\Omega$.
The dotted curves are calculated for the case 
$\bar \Omega= \bar \Omega_{eq}=1.55$. The dot-dashed curves are calculated for the equilibrium angular 
velocity of the planet, $\Omega_{ps}^{f}$, associated with the fundamental modes, see equation $(\ref {a 6})$ 
of Appendix. These curves give a minimal energy transfer $E_{f}^{min}$ as a function of $\Omega $, for a given
$\eta$.

As follows from Appendix, in order to calculate $E_{f}$ the 
eigenfrequencies associated with the fundamental modes, $\sigma_{f}$ 
as well as the corresponding dimensionless overlap integrals, $Q_{f}$ must be known. 
These quantities depend on the planet structure,
and therefore they are different, in principle, for planets with different masses and 
radii, see IP for their numerical values. 
In Figure \ref{Fig.16} we show the case of a planet with $M_{*}=M_{J}$ and $R_{*}=R_{J}$
(the upper dashed, dotted and dot-dashed curves), where we have $\sigma_{f}\approx 1.2\Omega_{*}$ and 
$Q_{f}\approx 0.56$. The lower dashed, dotted and dot-dashed curves represent the case of a planet with
the same value of mass and $R_{*}=2R_{J}$ with $\sigma_{f}\approx 1.7\Omega_{*}$ and 
$Q_{f}\approx 0.44$. As follows from the results provided in IP, when $M_{*}=5M_{J}$ 
the dependence of the eigenfrequencies and overlap integrals on the planet radius is rather
weak, and the corresponding energy transfer is always approximately represented by the
upper dashed and dotted curves in  Figure \ref{Fig.16}. 

\begin{figure}
\begin{center}
\vspace{7cm}\includegraphics{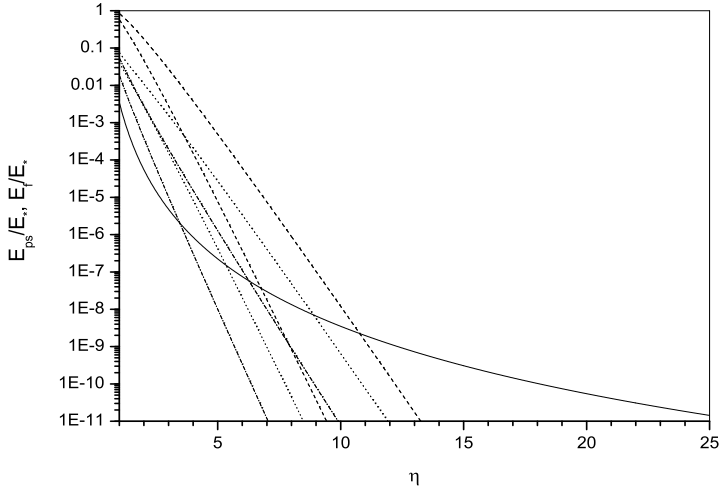}
\end{center}
\vspace{-1cm}
\caption{The energy transfer associated with the inertial modes (the solid curve)
as well as the energy transfer associated with the fundamental modes as a function
of the parameter $\eta $. The dashed, dotted and dot-dashed curves correspond to 
$\Omega =\Omega_{eq}$, $\Omega =0$ and $\Omega =\Omega_{ps}^{f}$, respectively.
Upper (lower ) curves of the same type 
corresponds to a planet with $M_{*}=M_{J}$ and $R_{*}=R_{J}$ ($R_{*}=2R_{J}$).}
\label{Fig.24}
\end{figure}

As follows from Figure \ref{Fig.16} the inertial modes dominate the energy transfer for sufficiently
large values of $\eta$. When $\bar \Omega =\bar \Omega_{eq}$ the contribution determined by the inertial
waves dominates over that corresponding to the fundamental modes for $\eta > 9$ in the case of $R_{*}=R_{J}$
and for $\eta > 5.5$ in the case of $R_{*}=2R_{J}$. Thus, we expect that the circularisation of the planet
orbits with sufficiently large periastron distances is mainly determined by the contribution associated with
the inertial modes.

\section{Time scale of orbital circularisation}

In this Section we apply the results developed in the previous Sections to estimate the effect of 
tides exerted by the central star on the planet on the planet orbital evolution. We assume that 
the state of pseudo-synchronisation with $\bar \Omega=\bar \Omega_{ps}\approx 1.55$
is maintained in a situation when the inertial modes dominate
over the fundamental modes in tidal response. 

Apart from the fact that we mainly consider the effects of inertial waves, the formulation of the problem
is analogous to what was discussed in IP. Let us consider a planetary system consisting of a few massive
planet and a central star and assume that a certain planet has lost a significant amount of its 
orbital angular momentum, most probably due to gravitational interaction with other planets in the
planetary system. Such situations have been frequently observed in numerical simulations of gravitationally 
interacting planets (Papaloizou $\&$ Terquem 2001, Adams $\&$ Laughlin 2003). The orbital semi-major 
axis of this planet, in this case, could be of the order of its 'typical' value $1-100Au$ while
the periastron distance $r_{p}$ could be very small, of the order of a few stellar radii. For a such
eccentric orbit the orbital angular momentum per unit of mass is related to the periastron distance as 
$L_{orb} \approx \sqrt {2GMr_{p}}$.

If we assume that the orbital evolution is determined by tides, the angular momentum is approximately
conserved while the semi-major axis may decrease due to transfer of the orbital energy into
oscillations of the planet by tides and subsequent dissipation. 
Then, the energy may be radiated away from the planet.
The conservation of the angular momentum leads to the well known fact that the radius a 'final' circular
orbit, $r_{f}$ is twice as large as the periastron distance $r_{p}$. Therefore, $r_{p}$ can be related 
to the observed orbital period of the planet after circularisation, $P_{obs}=2\pi \sqrt {(r^{3}_{f}/GM)}$, as   
\begin{equation}
r_{p}={r_{f} \over 2}=4.36(P_{3})^{2/3}\left({R_{\odot}\over R_{s}}\right)R_{s},
\label{eq p95}
\end{equation}  
where $P_{3}=P_{obs}/(3days)$, $R_{s}$ is the stellar radius 
and $R_{\odot}=7\cdot 10^{10}cm$ is the radius of the Sun. We have, accordingly, from equation 
$(\ref {eq p61})$
\begin{equation}   
\eta=9.1\eta_{0}\left({R_{\odot}\over R_{s}}\right )^{3/2}P_{3}, \quad 
\eta_{0}=\sqrt {{M_{*}\over M}{R^{3}_{s}\over R^{3}_{*}}}. 
\label{eq p96}
\end{equation}     
For the planet and the star with Jupiter and solar values of mass and radius, respectively,
$\eta_{0}\approx 1$. 

Let us consider the evolution of the semi-major axis of the eccentric planet orbit 
due to tidal interactions near the orbit periastron.
We can assume that the contributions from each of successive impulsive energy transfer can 
be added to each other only when one of two conditions is fulfilled: either the energy is dissipated
during two successive periastron passages or the condition for the stochastic instability
$(\ref {eq p78})$ is satisfied. In the latter case the evolution of the semi-major axis is stochastic,
and we assume below that, in this case, our evolution equations are written for 
averaged over some distributions values of the corresponding quantities.

Under these assumption we can write the evolution equation for semi-major axis, $a(t)$, 
in the form (IP)
\begin{equation} 
{\dot a_{10}\over a_{10}}=-{1\over t_{10}(t)\sqrt a_{10}},
\label{eq p97}
\end{equation}      
where $a_{10}=a/(10Au)$. The characteristic evolution time 
\begin{equation} 
t_{10}=7.4\cdot 10^{8}\left({M M_{J}\over M_{\odot}M_{*}}\right)
\left( {R_{*}\over R_{J}}\right)\left({10^{-9} \over \epsilon_{DT}}\right)  yr. 
\label{eq p98}
\end{equation}   
depends on the energy transfer due to dynamic tides associated with the inertial and fundamental
modes in the planet, $E_{ps}$ and $E_{f}$, respectively, 
and also on the transfer of the orbital energy into stellar oscillations
determined by dynamic tides exerted on the star 
\begin{equation}    
\epsilon_{DT}={(E_{ps}+E_{f}+E_{s})\over E_{*}},
\label{eq p99}
\end{equation}
where $E_{ps}$ is given by equation $(\ref {eq p94})$ and $E_{s}$ stands for the energy transfer
associated with dynamic tides in the star. Note that the time $t_{10}$ implicitly depends on 
time $t$ through dependence of the planet radius $R_{*}$ on time due to the evolutionary cooling and
a possible tidal heating of the planet. 

Equation  $(\ref {eq p98})$ can be integrated to give
\begin{equation}
a_{10}(t)=a_{in}\left(1-{1\over 2\sqrt a_{in}}\int_{t_{in}}^{t}
{dt^{'}\over t_{10}(t^{'})}\right )^{2},
\label{eq p100}
\end{equation}
where $a_{in}(t_{in})$ is the 'initial' value of the semi-major axis expressed in units of
$10Au$. A more accurate compared to  $(\ref {eq p98})$ 
definition of the evolution time scale, $t_{ev}$, can be obtained from equation $(\ref {eq p100})$ 
requiring that the semi-major axis $a_{10}$ is formally equal to zero when $t=t_{ev}$. Thus
\begin{equation}
{1\over 2\sqrt a_{in}}\int_{t_{in}}^{t_{ev}}{dt^{'}\over t_{10}(t^{'})}=1.
\label{eq p101}
\end{equation}

For a constant value of $t_{10}$, $t_{ev}\approx 2\sqrt a_{in} t_{10}$. In this case we 
can obtain a simple estimate of $t_{ev}$ assuming that the transfer of energy is determined only
by the inertial modes. Setting $E_{s}=E_{f}=0$ in equation $(\ref {eq p99})$ and 
substituting equations $(\ref {eq p94})$, $(\ref {eq p96})$ and
$(\ref {eq p99})$ into $(\ref {eq p98})$, we have
\begin{equation}
t_{ev}\sim 2\cdot 10^{8}\left({M_{*}\over M_{J}}\right )^{2}
\left ({M\over M_{\odot}}\right )^{-2}\left ({R_{*}\over 
R_{J}}\right)^{-8}P_{3}^{6}\sqrt{a_{in}} \hspace{4mm} yr. 
\label{eq p102}
\end{equation}
Note a very strong dependence of $t_{ev}$ on $R_{*}$ and $P_{orb}$. 

\begin{figure}
\begin{center}
\vspace{7cm}\includegraphics{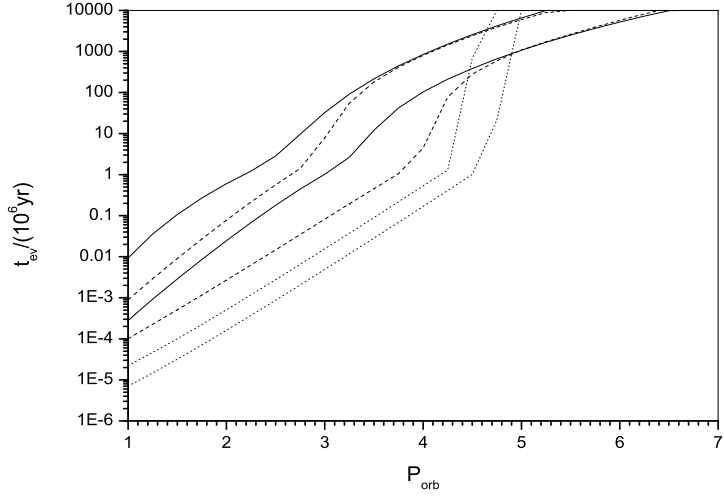}
\end{center}
\caption{The circularisation time scale $t_{ev}$ (in units of $10^{6}yr$)
for $1M_J$  as a function of the planet
orbital period after the stage of tidal circularisation, $P_{orb}$ (in units of days). The solid curves 
correspond to $t_{ev}$ determined solely by the inertial modes. The dashed curves are calculated 
with help of the sum of contributions corresponding to the inertial modes, the fundamental modes and the
modes excited in the star. The energy transfers 
determined by the inertial and fundamental modes are given by expressions
corresponding to rotation of the planet 
at the equilibrium angular velocities $\Omega_{ps}$ and $\Omega^{f}_{ps}$,
respectively. The dotted curves correspond to the case of a non-rotating planet. Accordingly, the contribution
of inertial modes is not taken into account. The lower (upper) curves of the same type correspond to 
$a_{in}= 1 \  [a= 10Au]$ ($a_{in}= 10 \  [a = 100Au]$).}
\label{Fig.25}
\end{figure}
\begin{figure}
\begin{center}
\vspace{7cm}\includegraphics{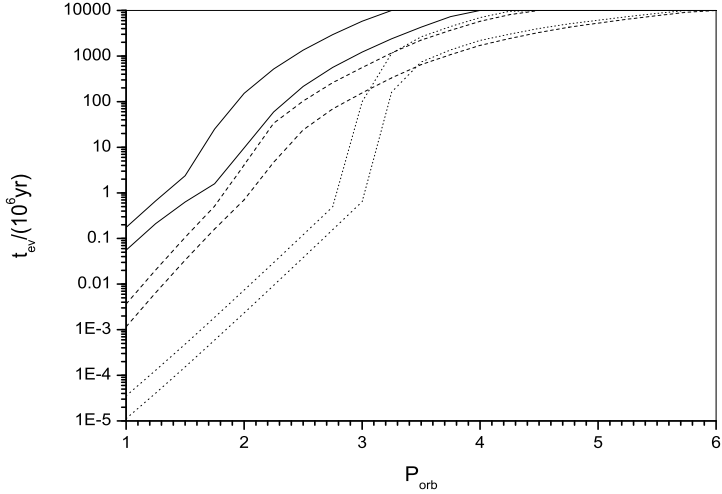}
\end{center}
\vspace{-1.cm}
\caption{The same as Figure \ref{Fig.25} but for $M_{*}=5M_{J}$.}
\label{Fig.26}
\end{figure}

Results of numerical calculations of $t_{ev}$ are shown in Figure \ref{Fig.25} for the case of
a planet with $M_{*}=1M_{J}$, and in  Figure \ref{Fig.26} for a planet with $M_{*}=5M_{J}$. In order
to calculate $t_{ev}$ we use expressions for the energy transfer associated with the fundamental
modes given in Appendix, the corresponding eigenfrequencies and overlap integrals are given in IP.
The dependence of $R_{*}$ on time is taken from 
Burrows et al (1997) where a cooling history of isolated planets is considered. Thus we neglect 
effects of tidal heating on the planet size and structure reserving their discussion 
for a future work. Burrows et al (1997) consider evolution of the planet radius starting from
the initial time $t_{in} \sim 10^{6}yr$ and until a maximal
time $t_{max}\sim 10^{9}yr$ is reached. We set the initial time $t_{in}=2\cdot 10^{6}yr$. 
We assume that the planet radius is not changed significantly
for $t > t_{max}$ and set $R_{*}(t > t_{max})=R_{*}(t_{max})$ with $R_{*}(t_{max})=7.2\cdot 10^{9}cm$
for $M_{*}=M_{J}$ and $R_{*}(t_{max})=7.67\cdot 10^{9}cm$ for $M_{*}=5M_{J}$. The lower curves of
the same type in Figures \ref{Fig.25} and \ref{Fig.26} correspond to $a_{in}=10Au$ and the upper
curves are calculated for $a_{in}=100Au$.

In Figure \ref{Fig.25} the solid curves describe the dependence of $t_{ev}$ on $P_{orb}$ calculated 
under assumption that only inertial waves contribute to the tidal evolution, and accordingly,
in this case $E_{st}=E_{f}=0$ in equation $(\ref {eq p99})$. For the dashed curves
we take into account all the contributions to the sum in $(\ref {eq p99})$
\footnote{Note that conditions for the  dissipation of the mode energy 
and the condition for the stochastic instability to set in are different 
for the inertial and fundamental modes in the planet and also for 
the modes determining the tidal response of the star.
Therefore, simple adding of different contributions to the sum in $(\ref {eq p99})$ is, strictly speaking,
not correct. However, we assume hereafter 
that this procedure gives correct order of magnitude estimates
reported in this Paper.}. In order to calculate $E_{s}$ we model the star as a non rotating
$n=3$ polytrope and take into account the contributions of $g$, $f$ and low order
$p$ modes, see IP for details. The contribution of the fundamental modes operating in the planet, $E_{f}$
is calculated 
for the case of rotation of the planet at the equilibrium angular velocity associated with
the fundamental modes, see equation $(\ref {a 6})$ of Appendix. 
In this case $E_{f}$ is given by equation $(\ref {a 7})$. This may be justified according to the following
arguments. The inertial modes dominates over the fundamental modes at sufficiently large values of $\eta$, 
and accordingly at sufficiently large values of $P_{orb}$, 
and the fundamental modes dominate at smaller values of
$P_{orb}$. Taking into account that the contribution of fundamental modes
decreases exponentially with increase of $P_{orb}$, the transition from the case of domination of 
the fundamental modes
to the case of domination of inertial modes occurs near a certain value of $P_{orb}$. Taking into
account that a planet with a low value of the moment of inertia has 
the angular velocity always close to
the equilibrium value determined by the modes dominating the energy transfer, 
the dashed curve could give an order of magnitude estimate of $t_{ev}$ for the whole range of $P_{orb}$.
The dotted curves show the results of calculation of $t_{ev}$ for the case of a non rotating planet.
In this case the energy transfer determined by the inertial modes is absent and we set $E_{ps}=0$ in
equation $(\ref {eq p99})$ and we use equation $(\ref {a 5})$ to calculate $E_{f}$. 

As seen from Figure \ref{Fig.25} when $M_{*}=1M_{J}$ the inertial 
modes dominate for $P_{orb} > 4-4.5days$.
The time scale of circularisation is smaller than a typical life time of a planetary system $\sim 5Gyr$
when $P_{orb} < 6days$ for $a_{in}=10Au$ and when $P_{orb} < 5days$ for $a_{in}=100Au$. It is interesting
to note that the contribution of the fundamental modes sharply decrease when
$t_{ev} > 10^{6} yr$. This is simply because the energy transfer associated with the 
fundamental modes is highly sensitive
to the radius of the planet (see Appendix). It can lead to a 
substantial change of the semi-major axis only when the planet radius is close to its 'initial value'
$R_{*}(t_{in})\sim 2R_{J}$. The cooling of the planet results in a  decrease of 
the planet radius on a time scale $\sim 10^{6}yr$ and the contribution of the fundamental modes
quickly become unimportant after this time is reached. The tides exerted on the
star are not significant at all for a planet with $M_{*}=1M_{J}$.

In Figure \ref{Fig.26} we show the results corresponding to a planet with $M_{*}=5M_{J}$. Contrary to the
previous case now the inertial modes are not important for the whole range of $P_{orb}$
and the circularisation time scale at large values of $P_{orb}$ is determined by tides exerted 
by the planet on the star. This may be explained as follows. At large values of $P_{orb}$ 
the circularisation time scale is of the order of several $Gyr$. A planet with its mass 
in the range $1M_{J}-5M_{J}$ and its age of the
order of a few $Gyr$ has the radius close to $R_{*}(t_{max})\sim 1R_{J}$. 
The circularisation time scale is mainly determined by the expressions for the energy 
transfer calculated for this radius.
Assuming that the inertial modes dominate we 
can use equation $(\ref {eq p102})$ to estimate $t_{ev}$. 
As follows from this equation 
$t_{ev}\propto M_{*}^{2}$. When eg. $P_{orb}=6days$ and $a_{in}=10Au$
we have $t_{ev}\sim 5Gyr$ for $M_{*}=1M_{J}$ and
$t_{ev}\sim 25Gyr$ for $M_{*}=5M_{J}$. Obviously, the inertial modes cannot 
lead to a significant decrease of the
semi-major distance in the latter case over the life time of the planetary systems. 
  Unlike the situation with the inertial modes,
 the energy transfer determined by the stellar modes 
grows with the planet mass, and 
therefore the tides exerted on the star dominate when the planet mass is sufficiently large.

\section{Conclusions and Discussion}
In this Paper we have developed and extended a new self-adjoint formalism 
for the problem of small stellar oscillations proposed in our recent paper(PI). 
We went on to apply this to the calculation of 
the tidal response of uniformly rotating fully convective planets and stars which
undergo a single encounter in a parabolic orbit or a sequence of 
multiple encounters in a weakly bound orbit. 
This was then applied to the problem of the tidal circularisation
of the orbits of the extrasolar planets. 

In section 2 we  began by  showing that the general formalism, presented in PI when
$W=\rho'c^2/\rho+\Psi^{int}$ is used as variable characterising linear eigenmodes, can also be applied  
when  the Lagrangian displacement is used. The advantage of this approach is that, as
no low frequency approximation is needed, it allows us to 
consider pulsations of  arbitrary frequency. We  
derived  general expressions for the energy and angular momentum transfer that occurs
when a rotating planet or a star passes through  periastron in a parabolic or highly eccentric 
orbit around a central mass 
The amplitudes and phases of the eigenmodes excited in  the planet or 
a star as a result of the tidal encounter were also determined.

We showed that these expressions can be 
represented in terms of a
 spectral decomposition over the normal modes of the self-adjoint operator
$(\ref {eq p15})$. 
We also obtained a self-consistent expression for the rate of dissipation of  an excited mode 
to leading order in the small parameter defined by the ratio of
pulsation period to a  characteristic  dissipation time scale. 

In section 3  we  focused on the case
of low frequency inertial modes, showing how the more general
formalism in terms of the Lagrangian displacement can be reduced to that in PI
in the low frequency limit.
This procedure 
 allows one to separate out the contribution associated with the inertial modes and simplify     
the equations governing  free pulsation of a rotating planet or a star. 
In that case the corresponding self-adjoint problem is formulated through
equations $(\ref {eq p47})$ and $(\ref {eq p48})$.
Note that the tidal response of the inertial modes to  impulsive tidal forcing
has not been described using other methods. 

We showed  in section 4 that when the rotating planet or a star approaches
periastron in an unperturbed state the tidal energy and angular momentum transfered through
the excitation of
inertial modes are given by very simple expressions $(\ref {eq p65})$ 
and $(\ref {eq p68})$.
 We considered the multi-passage problem assuming that there was no significant 
dissipation of  mode energy between successive periastron passages, and found a simple condition
$(\ref {eq p78})$ for the  occurrence of  stochastic instability that  results 
in the  stochastic gain of inertial mode energy over many periastron passages.
This was quantitatively similar to that given by IP for $f$ modes.
We found that  for  stochastic instability  the circularisation process has to start with 
 $a >  \sim 30 AU,$ for final periods of $\sim 3$ days.
or $\sim 1-2 AU$ for final orbital
periods $\sim 1.2$ days.

In section 5  we applied our formalism to  fully
convective rotating giant planets. 
In order to evaluate the expressions for the energy and angular momentum transfer
the eigenspectrum of  $(\ref {eq p48})$ must be found numerically, 
for a given model.
We calculated the eigenspectrum of the inertial modes for several
planet models with a realistic equation of state. 
The details and stability of the numerical 
method we used are discussed in section 5.2.
 We considered planet models 
with two different masses equal to $1M_{J}$ and $5M_{J}$, 
with  radii in the range $1-2R_{J}$. We found that,
as in the case of $n=1.5$ polytrope considered by PI,  the tidal response was determined by a few
'global' eigenmodes with a large scale distribution of  perturbed quantities. Two 'main' 
or 'standard'
global modes have eigenfrequencies close to those corresponding to the $n = 1.5$ polytrope.
These modes have the  largest overlap integrals characterising coupling between the tidal
field and the eigenmodes.
However, we found that for realistic planet models,
there could be  'non-standard' global modes possibly related to a sharp change 
of structure occurring  near to the point of ionisation of hydrogen. The structure of 
the 'non-standard' modes as well as their stability with respect to a change 
of the planet mass and radius was  discussed. 

The results obtained for the eigen spectra of realistic planet
models were applied to the problem of tidal circularisation of  
the orbits of the extrasolar planets in sections 6 and 7.
In section 6 we calculated the rotational angular velocity
for which the net angular momentum transferred was zero.
Because of the relative low inertia of the planet compared to the orbit
it is expected to rapidly attain this angular velocity
and achieve a state of pseudo synchronisation. The angular velocity 
associated with pseudo synchronisation was found to be always
close to $1.55$ times the angular velocity  of a circular orbit at periastron.

Orbital evolution arises  through  the transfer of  orbital energy to the 
modes of oscillation. Transfers at successive periastron passages
lead to orbital circularisation. This is the case,
either when the mode energy is dissipated directly between encounters,
or when the system is in the regime of  stochastic instability.
We compared the contribution   associated with the inertial modes 
with the
contribution associated with the fundamental modes, in a state of pseudo synchronisation,
to the circularisation time scale in section 7.
We found  that the inertial modes  
led to effective circularisation of a giant planet with mass $\sim 1M_{J}$ on a time scale smaller
than or of the order of a few $Gyr$, when  the initial semi major axis
was less than $10 AU$ and the final orbital period after circularisation 
$P_{orb} < 6days.$ In this case the inertial modes  play the most important role
at the longer periods. However, the inertial waves are less important for the planets
 of  larger mass $5M_{J}$ that we considered.
In that case the dynamic tides exerted on the star central star, which we assumed to be
solar like,  play
a major role.

In our opinion, the most important unresolved question related to the theory discussed above
is concerned with internal  dissipation. Of particular
concern is the internal location of modal dissipation and the related issue of 
 whether the energy deposited
can be  radiated away between successive periastron passages.
 There are several potentially interesting channels of
dissipation which may operate in our case. The non-linear parametric instability 
discussed by Kumar and Goodman (1996) for the case of $g$ modes in convectively stable stars 
may be effective for the inertial modes as well. This instability operates when
there are eigenmodes in the stars with eigenfrequencies approximately half  that of
the eigenfrequencies of the tidally excited modes and sufficiently large coupling coefficients.
These modes may be unstable on account of parametric resonance.
Taking into account that the eigenspectrum of the inertial modes is dense, 
 unstable  modes are expected to  exist for sufficiently  
weak dissipative processes. 
This question will be addressed in our future work.

Our formalism is general enough that it  can be extended to the case of convectively stable uniformly rotating stars and it can
 also be applied to other astrophysical situations where excitation of stellar
pulsations with frequencies comparable 
to the angular velocity of rotation is important.
In this particular work we assumed that the orbital 
and planetary angular momentum vectors were aligned.
This is a very reasonable assumption when the system has many
periastron passages and so readily attains a state of pseudo
synchronisation during the circularisation process.
However, misaligned angular momenta should be considered
if the main issue is tidal capture, as in globular clusters,
because in that case there is no reason for the initial
angular momentum vectors to be correlated.   
These issues and extensions of the formalism
  will  be considered in future work.

\vfill
\eject

\vspace{-0.7cm}
  
\section*{Acknowledgements}

\vspace{-0.1cm}

We are grateful to G. Ogilvie for discussions.
PBI has been supported in part
by RFBR grant 04-02-17444.

\vspace{-0.4cm}

\appendix
\section{Energy and angular momentum transfer associated with the fundamental modes}

Here we show the expressions for energy and angular momentum transfer associated
with the the fundamental modes in the limit $\eta \rightarrow \infty$ (IP). For the energy transfer
we have
\begin{equation}
E_{f}={16\sqrt 2 \over 15}\pi^{2}\tilde \sigma_{f}^{3} Q_{f}^{2}\eta 
e^{-{4\sqrt 2 \over 3}\tilde \sigma_{+} \eta}F_{1}(\tilde \Omega )E_{*},
\label{a 1}
\end{equation}
and for the angular momentum transfer we have
\begin{equation}
L_{f}={32\sqrt 2 \over 15}\pi^{2}\tilde \sigma_{f}^{2}Q_{f}^{2}\eta 
e^{-{4\sqrt 2 \over 3}\tilde \sigma_{+} \eta}F_{2}(\tilde \Omega )L_{*},
\label{a 2}
\end{equation}
where 
\begin{equation}
F_{1}(\tilde \Omega )=
1+{3\over 2^{6}(\tilde \sigma_{f}\eta)^{2}}e^{{8\sqrt 2\over 3}\beta \tilde \Omega \eta}+
{9\over 2^{14}(\tilde \sigma_{f}\eta)^{4}}e^{{16\sqrt 2\over 3}\beta \tilde \Omega \eta}
\label{a 3}
\end{equation}
and
\begin{equation}
F_{2}(\tilde \Omega )=
1-{9\over 2^{14}(\tilde \sigma_{f}\eta)^{4}}e^{{16\sqrt 2\over 3}\beta \tilde \Omega \eta}.
\label{a 4}
\end{equation}
$\tilde \sigma_{f}=\sigma_{f}/\Omega_{*}$ and $\tilde \Omega =\Omega /\Omega_{*}$ 
are the dimensionless eigenfrequency and dimensionless angular velocity of the planet, respectively
and $\tilde \omega_{+}=\tilde \sigma_{f} + 2\beta \tilde \Omega$. The dimensionless quantity $\beta $
determines the frequency splitting of the mode due to rotation, it is close to $0.5$ for our models 
(see IP and references therein). $Q_{f}$ are the dimensionless overlap integrals, see PT and IP.
When $\Omega = 0$ the energy transfer as a function of $\tilde \Omega $ is maximal,
\begin{equation}
E_{f}^{max}\approx {16\sqrt 2 \over 15}\pi^{2}\tilde \sigma_{f}^{3} Q_{f}^{2}\eta 
e^{-{4\sqrt 2 \over 3}\tilde \sigma_{f} \eta}E_{*}.  
\label{a 5}
\end{equation}
Assuming that the fundamental modes dominate the tidal response, the state of pseudo-synchronisation
is determined by condition $L_{f}(\tilde \Omega_{ps}^{f})=0$. In this case we obtain from equation
$(\ref{a 4})$
\begin{equation}
\Omega_{ps}^{f}={3 \over 4\sqrt 2 \beta \eta}\ln {(\sqrt {2\over 3} 8\sigma_{f}\eta)}.
\label{a 6}
\end{equation}
When $\tilde \Omega = \tilde \Omega_{ps}^{f}$ the energy transfer is minimal  
\begin{equation}
E_{f}^{min}\approx {\pi^{2}\over 5\sqrt 2}{Q_{f}^{2}\over \eta}
e^{-{4\sqrt 2 \over 3}\tilde \sigma_{f} \eta}E_{*}.
\label{a 7}
\end{equation}

\bsp

\label{lastpage}

\end{document}